\documentclass[aoas,preprint]{imsart}
\RequirePackage[OT1]{fontenc}
\RequirePackage{amsthm,amsmath}
\RequirePackage[authoryear]{natbib}
\RequirePackage{graphicx}
\usepackage{pgf, tikz,float}
\usetikzlibrary{arrows,shapes,positioning}
\usetikzlibrary{calc,decorations.markings,automata}
\usepackage{multirow}
\usepackage{comment}
\RequirePackage[colorlinks,citecolor=blue,urlcolor=blue]{hyperref}
\usepackage{subcaption}
\usepackage{xcolor}

\newtheorem{assumption}{Assumption}
\newtheorem{theorem}{Theorem}

\newcommand\independent{\protect\mathpalette{\protect\independenT}{\perp}}
\def\independenT#1#2{\mathrel{\rlap{$#1#2$}\mkern2mu{#1#2}}}

\newcommand{\bM}{\mathbf{M}}
\newcommand{\bY}{\mathbf{Y}}

\newcommand{\ACME}{\mbox{\tiny{ACME}}}
\newcommand{\ANDE}{\mbox{\tiny{ANDE}}}
\newcommand{\TE}{\mbox{\tiny{TE}}}
\newcommand{\RD}{\mbox{\tiny{RD}}}
\newcommand{\GEE}{\mbox{\tiny{GEE}}}

\newcommand{\one}{\mathbf{1}}

\startlocaldefs
\numberwithin{equation}{section}
\theoremstyle{plain}

\endlocaldefs

\begin{document}
	
	\begin{frontmatter}
		\title{Causal Mediation Analysis for Sparse and Irregular Longitudinal Data}
		\runtitle{Mediation Analysis with Longitudinal Data}

		\begin{aug}
			\author[A]{\fnms{Shuxi} \snm{Zeng}\ead[label=e1,mark]{ zengshx777@gmail.com}},
			\author[B,D]{\fnms{Stacy} \snm{Rosenbaum}\ead[label=e2]{rosenbas@umich.edu}},
			\author[C]{\fnms{Susan} \snm{C. Alberts}\ead[label=e3]{alberts@duke.edu}},
			\author[D]{\fnms{Elizabeth} \snm{A. Archie}\ead[label=e4]{earchie@nd.edu}}
			\and
			\author[A]{\fnms{Fan} \snm{Li}\ead[label=e5,mark]{fl35@duke.edu}}
			\address[A]{Department of Statistical Science, Duke University, \printead{e1,e5}}
			\address[B]{Department of Anthropology, University of Michigan, \printead{e2}}
			\address[C]{Departments of Biology and Evolutionary Anthropology, Duke University, \printead{e3}}
			\address[D]{Department of Biological Sciences, University of Notre Dame, \printead{e4}}
		\end{aug}
		
		\begin{abstract}
			Causal mediation analysis seeks to investigate how the treatment effect of an exposure on outcomes is mediated through intermediate variables. Although many applications involve longitudinal data, the existing methods are not directly applicable to settings where  the mediator and outcome are measured on sparse and irregular time grids. 
			We extend the existing causal mediation framework from a functional data analysis perspective, viewing the sparse and irregular longitudinal data as realizations of underlying smooth stochastic processes. We define causal estimands of direct and indirect effects accordingly and provide corresponding identification assumptions. For estimation and inference, we employ a functional principal component analysis approach for dimension reduction and use the first few functional principal components instead of the whole trajectories in the structural equation models. We adopt the Bayesian paradigm to accurately quantify the uncertainties. The operating characteristics of the proposed methods are examined via simulations. We apply the proposed methods to a longitudinal data set from a wild baboon population in Kenya to investigate the causal relationships between early adversity, strength of social bonds between animals, and adult glucocorticoid hormone concentrations. We find that early adversity has a significant direct effect (a 9-14\% increase) on females' glucocorticoid concentrations across adulthood, but find little evidence that these effects were mediated by weak social bonds. 
		\end{abstract}
		
		
	\end{frontmatter}

	\section{Introduction\label{Sec_Intro}}
	
	
	Mediation analysis seeks to understand the role of an intermediate variable (i.e. mediator) $M$ that lies on the causal path between an exposure or treatment $Z$ and an outcome $Y$. The most widely used mediation analysis method, proposed by \cite{baron1986moderator}, fits two linear structural equation models (SEMs) between the three variables and interprets the model coefficients as causal effects. There is a vast literature on the Baron-Kenny framework across a variety of disciplines, including psychology, sociology, and epidemiology (\citealp[see][]{mackinnon2012introduction}). A major advancement in recent years is the incorporation of the potential-outcome-based causal inference approach \citep{Neyman1923, Rubin1974}. This led to a formal definition of relevant causal estimands, clarification of identification assumptions, and new estimation strategies beyond linear SEMs \citep{robins1992identifiability,pearl2001direct, sobel2008identification,tchetgen2012semiparametric,daniels2012bayesian, vanderweele2016mediation}. In particular, \cite{imai2010identification} proved that the Baron-Kenny estimator can be interpreted as a special case of a causal mediation estimator given additional assumptions. 
	These methodological advancements opened up new application areas including imaging, neuroscience, and environmental health \citep{lindquist2011graphical, lindquist2012functional,zigler2012estimating, kim2019bayesian}. Comprehensive reviews on causal mediation analysis are given in \cite{vanderweele2015explanation,nguyen2019clarifying}.
	
	In the traditional settings of mediation analysis, exposure $Z$, mediation $M$ and outcome $Y$ are all univariate variables at a single time point. Recent work has extended to time-varying cases, where at least one of the triplet $(Z, M, Y)$ is longitudinal. This line of research has primarily focused on cases with time-varying mediators or outcomes that are observed on sparse and regular time grids \citep{van2008direct,roth2012mediation,lin2017parametric}. For example, \cite{vanderweele2017timevarying} developed a method for identifying and estimating causal mediation effects with time-varying exposures and mediators based on marginal structural models \citep{robins2000msm}. Some researchers also investigated the case with time-varying exposure and mediator for the survival outcome (\cite{zheng2017longitudinalsurvival,lin2017mediationsurvival}). Another stream of research, motivated by applications in neuroimaging, focuses on cases where mediators or outcomes are densely recorded continuous functions, e.g. the blood-oxygen-level-dependent (BOLD) signal collected in a functional magnetic resonance imaging (fMRI) session. In particular, \cite{lindquist2012functional} introduced the concept of \emph{functional mediation} in the presence of a functional mediator and extended causal SEMs to functional data analysis \citep{ramsay2005functional}. \cite{zhao2018functional} further extended this approach to functional exposure, mediator and outcome.
	
	Sparse and irregularly-spaced longitudinal data are increasingly available for causal studies. For example, in electronic health records (EHR) data, the number of observations usually varies between patients and the time grids are uneven. The same situation applies in animal behavior studies due to the inherent difficulties in observing wild animals. Such data structure poses challenges to existing causal mediation methods. First, one cannot simply treat the trajectories of mediators and outcomes as functions as in \cite{lindquist2012functional} because the sparse observations render the trajectories volatile and non-smooth. Second, with irregular time grids the dependence between consecutive observations changes over time, making the methods based on sparse and regular longitudinal data such as \cite{vanderweele2017timevarying} not applicable. A further complication arises when the mediator and outcome are measured with different frequencies even within the same individual.
	
	In this paper, we propose a causal mediation analysis method for sparse and irregular longitudinal data that address the aforementioned challenges. Similar to \cite{lindquist2012functional} and \cite{zhao2018functional}, we adopt a functional data analysis perspective \citep{ramsay2005functional}, viewing the sparse and irregular longitudinal data as realizations of underlying smooth stochastic processes. We define causal estimands of direct and indirect effects accordingly and provide assumptions for nonparametric identification (Section \ref{Sec_Framework}). For estimation and inference, we proceed under the classical two-SEM mediation framework \citep{imai2010identification} but diverge from the existing methods in modeling (Section \ref{Sec_Modeling}). Specifically, we employ the functional principal component analysis (FPCA) approach \citep{yao2005functional,jiang2010covariatefpca,jiang2011functional,han2018functional} to project the mediator and outcome trajectories to a low-dimensional representation. We then use the first few functional principal components (instead of the whole trajectories) as predictors in the structural equation models. To accurately quantify the uncertainties, we employ a Bayesian FPCA model \citep{kowal2020bayesian} to simultaneously estimate the functional principal components and the structural equation models. Though the Bayesian approach to mediation analysis has been discussed before \citep{daniels2012bayesian,kim2017framework,kim2018bayesian}, it has not been developed for the setting of sparse and irregular longitudinal data.

	Our motivating application is the evaluation of the causal relationships between early adversity, social bonds, and physiological stress in wild baboons (Section \ref{sec:background}). Here the exposure is early adversity (e.g. drought, maternal death before reaching maturity), the mediators are the strength of adult social bonds, and the outcomes are adult glucocorticoid (GC) hormone concentrations, which is a measure of an animal's physiological stress level. The exposure, early adversity, is a binary variable measured at one time point, whereas both the mediators and outcomes are sparse and irregular longitudinal variables. We apply the proposed method to a prospective and longitudinal observational data set from the Amboseli Baboon Research Project located in the Amboseli ecosystem, Kenya \citep{alberts2012amboseli} (Section \ref{Sec_Application}). We find that experiencing one or more sources of early adversity leads to significant direct effects (a 9-14\% increase) on females’ GC concentrations across adulthood, but find little evidence that these effects were mediated by weak social bonds. Though motivated from a specific application, the proposed method is readily applicable to other causal mediation studies with similar data structure, including EHR and ecology studies. Furthermore, our method is also applicable to regularly spaced longitudinal observations.

	\section{Motivating Application: Early Adversity, Social Bond and Stress} \label{sec:background}
	\subsection{Biological Background} 
	Conditions in early life can have profound consequences for individual development, behavior, and physiology across the life course \citep{lindstrom1999early,gluckman2008effect,bateson2004developmental}. These early life effects are important, in part, because they have major implications for human health. One leading explanation for how early life environments affect adult health is provided by the biological embedding hypothesis, which posits that early life stress causes developmental changes that create a “pro-inflammatory” phenotype and elevated risk for several diseases of aging \citep{miller2011psychological}. The biological embedding hypothesis proposes at least two, non-exclusive causal pathways that connect early adversity to poor health in adulthood. In the first pathway, early adversity leads to altered hormonal profiles that contribute to downstream inflammation and disease. Under this scenario, stress in early life leads to dysregulation of hormonal signals in the body’s main stress response system, leading to the release of GC hormone, which engages the body’s fight-or-flight response. Chronic activation is associated with inflammation and elevated disease risk \citep{mcewen1998stress,miller2002chronic,mcewen2008central}. In the second causal pathway, early adversity hampers an individual’s ability to form strong interpersonal relationships. Under this scenario, the social isolation contributes to both altered GC profiles and inflammation. 
	
	Hence, the biological embedding hypothesis posits that early life adversity affects both GC profiles and social relationships in adulthood, and that poor social relationships partly mediate the connection between early adversity and GCs. Importantly, the second causal pathway—mediated through adult social relationships—suggests an opportunity to transmit the negative health effect of early adversity. Specifically, strong and supportive social relationships may dampen the stress response or reduce individual exposure to stressful events, which in turn reduces GCs and inflammation. For example, strong and supportive social relationships have repeatedly been linked to reduced morbidity and mortality in humans and other social animals \citep{holt2010social,silk2007adaptive}. In addition to the biological embedding hypothesis, this idea of social mediation is central to several hypotheses that propose causal connections between adult social relationships and adult health, even independent of early life adversity; these hypotheses include the stress buffering and stress prevention hypotheses \citep{cohen1985stress,landerman1989alternative,thorsteinsson1999meta} and the social causation hypothesis \citep{marmot1991health,anderson2011effects}.

	Despite the aforementioned research, the causal relationships among early adversity, adult social relationships, and HPA (hypothalamic–pituitary–adrenal) axis dysregulation remain the subject of considerable debate. While social relationships might exert direct effects on stress and health, it is also possible that poor health and high stress limit an individual’s ability to form strong and supportive relationships. As such, the causal arrow flows backwards, from stress to social relationships  \citep{case2011long}. 
	In another causal scenario, early adversity exerts independent effects on social relationships and the HPA axis, and correlations between social relationships and GCs are spurious, arising solely as a result of their independent links to early adversity \citep{marmot1991health}. 

	\subsection{The Data}
	\label{sec:data_general}
	In this paper, we test whether the links between early adversity, the strength of adult social bonds, and GCs are consistent with predictions derived from the biological embedding hypothesis and other related theories. Specifically, we use data from a well-studied population of savannah baboons in the Amboseli ecosystem in Kenya. Founded in 1971, the Amboseli Baboon Research Project has prospective longitudinal data on early life experiences, and fine-grained longitudinal data on adult social bonds and GC hormone concentrations, a measure of the physiological stress response \citep{alberts2012amboseli}. 
	
	Our study sample includes 192 female baboons. Each baboon entered the study after becoming mature at age 5, and we had information on its experience of six sources of early adversity (i.e., exposure) \citep{tung2016cumulative, zipple2019intergenerational}: drought, maternal death, competing sibling, high group density, low maternal rank, and maternal social isolation. Table \ref{tab:adversity} presents the number of baboons that experienced each early adversity. Overall, while only a small proportion of subjects experienced any given source of early adversity, most subjects experienced at least one source of early adversity. Therefore, in our analysis we also create a cumulative exposure variable that summarizes whether a baboon experienced any source of the adversity.
	\begin{table}[!ht]
		\centering
		\caption{Sources of early adversity and the number of baboons experienced each type of early adversity. The last row summarizes the number of baboons had at least one of six individual adversity sources.}  \label{tab:adversity}
		\begin{tabular}{l cc}
			\hline
			\hline
			early adversity& no. subjects did not experience & no. subjects did experience
			\\
			&(control) & (exposure) \\
			\hline
			Drought &164 &28\\
			Competing Sibling &153& 39\\
			High group density &161& 31\\
			Maternal death & 157 &35\\
			Low maternal rank & 152& 40\\
			Maternal Social isolation & 140& 52\\
			At least one & 48 &144\\
			\hline
		\end{tabular}
	\end{table}
	
	Each baboon's adult social bonds (i.e. mediators) and fecal GC hormone concentrations (i.e. outcomes) are measured repeatedly throughout its life on the same grid.  Social bonds are measured using the dyadic sociality index with females (DSI-F) \citep{silk2006social}. The indices are calculated for each female baboon separately based on all visible observations for social interactions between the baboon and other members in the entire social group within a given period. Larger values mean stronger social bonds. We 
	normalized the DSI-F measurements,
	and the normalized DSI-F values range from $-1.47$ to $3.31$ with mean value at $1.04$ and standard deviation $0.51$. The fecal GC concentrations were collected opportunistically, and the values range from $7.51$ to $982.87$ with mean $74.13$ and standard deviation $38.25$. Age is used to index within-individual observations on both social bond and GC concentrations. Only about 20\% baboons survive until age 18 and thus data on females older than 18 years are extremely sparse and volatile. Therefore, we truncated all trajectories at age 18, resulting in a final sample with 192 female baboons and 9878 observations. 
	
\begin{figure}[h]
		\centering
		\includegraphics[width=0.8\textwidth]{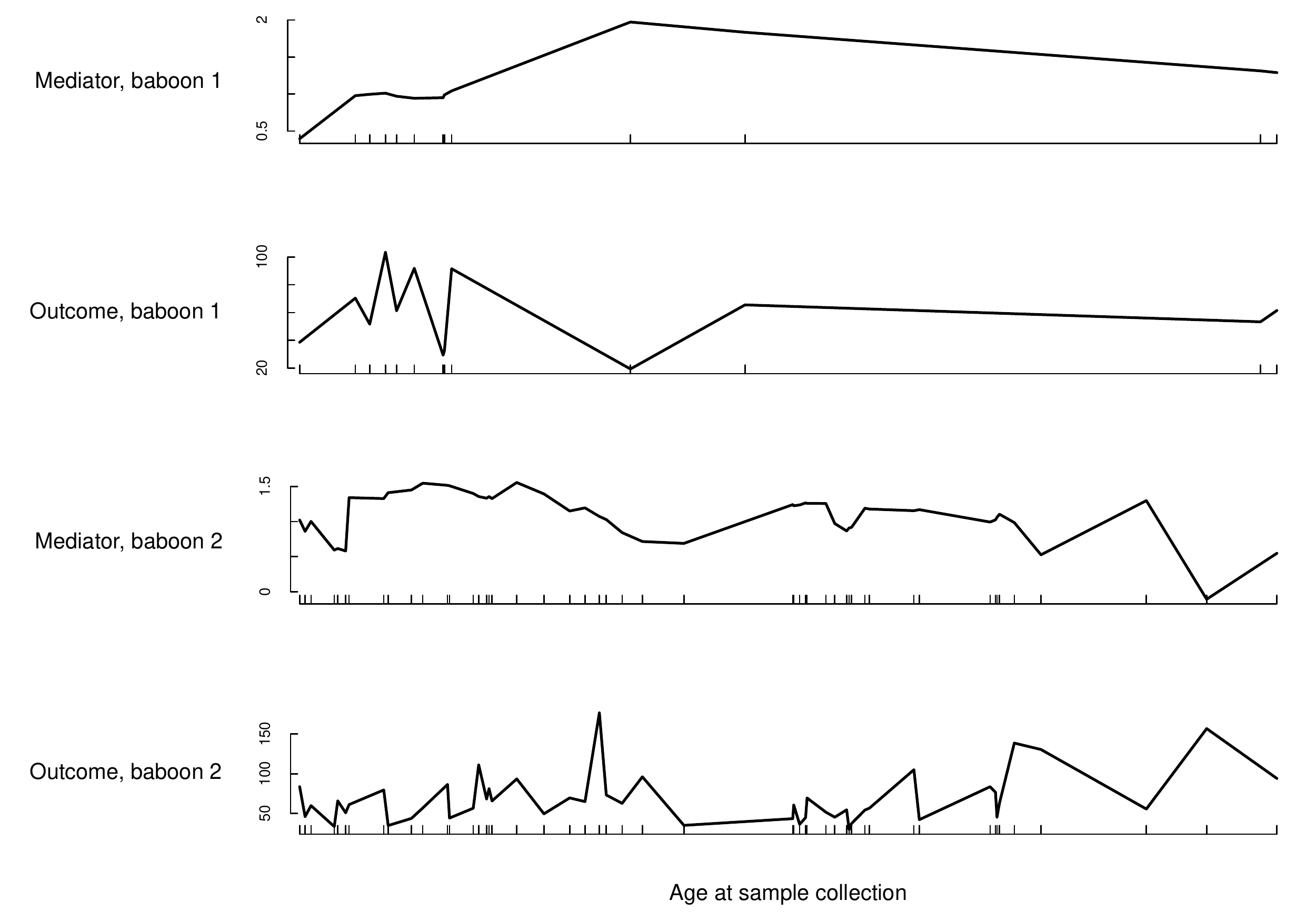}
		\caption{\label{fig:trajectories} Observed trajectories of social bonds and GC hormone as a function of age of two randomly selected female baboons in the study sample.}
	\end{figure}

	For wild animals, the observations usually made on irregular or opportunistic basis. We have on average 51.4 observations of each baboon for both social bonds and GC concentrations, but the number of observations of a single baboon ranges from 3 to 113. Figure \ref{fig:trajectories} shows the mediator and outcome trajectories as a function of age of two randomly selected baboons in the sample. We can see that the frequency of the observations and time grids of the mediator or outcome trajectories vary significantly between baboons. 
	\vspace{-1em}
	
	
	We also have a set of static and time-varying covariates that are deemed important to wild baboons's physiology and behavior. These include reproductive state (i.e. cycling, pregnant, or lactating), density of the social group, max temperature in the last 30 days before the fecal sample was collected, whether the sample is collected in wet or dry season, the amount of rainfall, relative dominance rank of a baboon, and number of coresident
	adult maternal relatives.More information on the covariates, exposure, mediator, and outcomes can be found in  \cite{rosenbaum2020pnas}. 
	
	\section{Causal Mediation Framework} \label{Sec_Framework}
	\subsection{Setup and Causal Estimands} \label{Sec_Setup}
	Suppose we have a sample of $N$ units (in the use case described here, baboons); each unit  $i\ (i=1,2,\cdots,N)$ is assigned to a treatment ($Z_{i}=1$) or a control ($Z_{i}=0$) group. For each unit $i$, we make observations at $T_{i}$ different time points $\{t_{ij}\in [0,T], j=1,2,\cdots,T_{i}\}$, and $T_i$ can vary between units. At each time point $t_{ij}$, we measure an outcome $Y_{ij}$ and a mediator $M_{ij}$ prior to the outcome, and a vector of $p$ time-varying covariates $\mathbf{X}_{ij}=(X_{ij,1},\cdots,X_{ij,p})'$. For each unit, the observations points are sparse along the time span and irregularly spaced. For simplicity, we assume the observed time grids for the outcome and the mediator are the same within one unit. However, our method is directly applicable when the observation grids for the outcome and the mediator are different for a given individual.
	
	A key to our method is to view the observed mediator and outcome values drawn from a smooth underlying process $M_{i}(t)$ and $Y_{i}(t)$, $t\in[0,T]$, with Normal measurement errors, respectively:
	\begin{eqnarray}
		M_{ij}&=&M_{i}(t_{ij})+\varepsilon_{ij}, \quad \varepsilon_{ij}\sim \mathcal{N}(0,\sigma_{m}^{2}),\\
		Y_{ij}&=&Y_{i}(t_{ij})+\nu_{ij},\quad \nu_{ij}\sim\mathcal{N}(0,\sigma_{y}^{2}).
	\end{eqnarray}
	Hence, instead of directly exploring the relationship between the treatment $Z_{i}$, mediators $M_{ij}$ and outcomes $Y_{ij}$, we investigate the relationship between $Z_{i}$ and the stochastic processes $M_{i}(t_{ij})$ and $Y_{i}(t_{ij})$. In particular, we wish to answer two questions: (a) how big is the causal impact of the treatment on the outcome process, and (b) how much of that impact is mediated through the mediator process? 
	
	To be consistent with the standard notation of potential outcomes in causal inference \citep{ImbensRubin2015}, from now on we move the time index of the mediator and outcome process to the superscript: $M_{i}(t)=M_{i}^{t},Y_{i}(t)=Y_{i}^{t}$. Also, we use the following bold font notation to represent a process until time $t$: $\mathbf{M}_{i}^{t}\equiv \{M_{i}^{s},s\leq t\}\in \mathcal{R}^{[0,t]}$, and  $\mathbf{Y}_{i}^{t}\equiv\{Y_{i}^{s},s\leq t\} \in \mathcal{R}^{[0,t]}$. Similarly, we denote covariates between the $j$th and $j+1$th time point for unit $i$ as  $\mathbf{X}_{i}^{t}=\{X_{i1},X_{i2},\cdots,X_{ij'}\}$ for $t_{ij'}\leq t<t_{ij'+1}$. 
	
	We extend the definition of potential outcomes to define the causal estimands. Specifically, let $\mathbf{M}_{i}^{t}(z)\in \mathcal{R}^{[0,t]}$ for $z=0,1,t\in[0,T]$, denote the potential values of the underlying mediator process for unit $i$ until time $t$ under the treatment status $z$; let $\mathbf{Y}_{i}^{t}(z,\mathbf{m})\in\mathcal{R}^{[0,t]}$ be the potential outcome for unit $i$  until time $t$ under the treatment status  $z$ and the mediator process taking value of $\mathbf{M}_{i}^{t}=\mathbf{m}$ with $\mathbf{m}\in \mathcal{R}^{[0,t]}$. 
	The above notation implicitly makes the stable unit treatment value assumption (SUTVA) \citep{rubin1980randomization}, which states that (i) there is no different version of the treatment, and (ii) there is no interference between the units, more specifically, the potential outcomes of one unit do not depend on the treatment and mediator values of other units. SUTVA is plausible in our application. First, there is unlikely different versions of the early adversities. Second, though baboons live in social groups, it is unlikely a baboon's long-term GC concentration (outcome) was much affected by the early adversities experienced by other cohabitant baboons in its social group, particularly considering the fact that only a small proportion of baboons experienced any given early adversity. Moreover, the social bond index (mediator) summarizes the interaction between a focal baboon and other members in a social group, and thus we can view the impact from other baboons as constant while examining the variation of social bond for the focal baboon. The notation of $\mathbf{Y}_{i}^{t}(z,\mathbf{m})$ makes another implicit assumption that the potential outcomes are determined by the mediator values $\mathbf{m}$ before time $t$, but not after $t$. For each unit, we can only observe one realization from the potential mediator or outcome process: 
	\begin{eqnarray}
		&&\mathbf{M}_{i}^{t}=\mathbf{M}_{i}^{t}(Z_{i})=Z_{i}\bM_{i}^{t}(1)+(1-Z_{i})\bM_{i}^{t}(0),\\
		&&\mathbf{Y}_{i}^{t}=\mathbf{Y}_{i}^{t}(z,\mathbf{M}_{i}^{t}(Z_{i}))=Z_{i}\bY_{i}^{t}(1,\mathbf{M}_{i}^{t}(1))+(1-Z_{i})\bY_{i}^{t}(0,\mathbf{M}_{i}^{t}(0)).
	\end{eqnarray}
	
	We define the total effect (TE) of the treatment $Z_{i}$ on the outcome process at time $t$ as:
	\begin{eqnarray}
		\label{Definition_ATE_Process}
		\tau_{\TE}^{t}&=&E\{Y_{i}^{t}(1,\mathbf{M}_{i}^{t}(1))-Y_{i}^{t}(0,\mathbf{M}_{i}^{t}(0))\}.
	\end{eqnarray}
	When there is a mediator, the TE can be decomposed into direct and indirect effects. Below we extend the framework of \cite{imai2010identification} to formally define these effects. First, we define the average causal mediation (or indirect) effect (ACME) under treatment $z$ at time $t$ by fixing the treatment status while altering the mediator process: 
	\begin{eqnarray}
		\label{Definition_ACME_Process}
		\tau_{\ACME}^{t}(z)&\equiv&E\{ Y_{i}^{t}(z,\mathbf{M}_{i}^{t}(1))- Y_{i}^{t}(z,\mathbf{M}_{i}^{t}(0))\},\quad z=0,1.
	\end{eqnarray}
	The ACME quantifies the difference between the potential outcomes, given a fixed treatment status $z$, corresponding to the potential mediator process under treatment $\mathbf{M}_{i}^{t}(1)$ and that under control $\mathbf{M}_{i}^{t}(0)$. In the previous literature, variants of the ACME are also called the \emph{natural indirect effect} \citep{pearl2001direct}, or the \emph{pure indirect effect} for $\tau_{\ACME}^{t}(0)$ and \emph{total indirect effect} for $\tau_{\ACME}^{t}(1)$ \citep{robins1992identifiability} 
	
	%
	Second, we define the average natural direct effect (ANDE) \citep{pearl2001direct,imai2010identification} of treatment on the outcome at time $t$ by fixing the mediator process while altering the treatment status:
	\begin{eqnarray}
		\label{Definition_ANDE_Process}
		\tau_{\ANDE}^{t}(z)&\equiv&E\{ Y_{i}^{t}(1,\mathbf{M}_{i}^{t}(z))-Y_{i}^{t}(0,\mathbf{M}_{i}^{t}(z))\}.
	\end{eqnarray}
	The ANDE quantifies the portion in the TE that does not pass through the mediators. 
	
	It is easy to verify that the TE is the sum of ACME and ANDE:
	\begin{eqnarray}
		\label{TE_decomposition}
		\tau_{\TE}^{t}=\tau_{\ACME}^{t}(z)+\tau_{\ANDE}^{t}(1-z), \quad z=0,1.
	\end{eqnarray}
	This implies we only need to identify two of the three quantities $\tau_{\TE}$, $\tau_{\ACME}^{t}(z)$, $\tau_{\ANDE}^{t}(z)$. In this paper, we will focus on the estimation of $\tau_{\TE}$ and $\tau_{\ACME}^{t}(z)$. Because we only observe a portion of all the potential outcomes, we cannot directly identify these estimands from the observed data, which would require additional assumptions. 

	\subsection{Identification assumptions}\label{Sec_Assumption}
	In this subsection, we list the causal assumptions necessary for identifying the ACME and ANDEs with sparse and irregular longitudinal data. There are several sets of identification assumptions in the literature \citep{robins1992identifiability,pearl2001direct,imai2010general,shpitser2011complete} with subtle distinction \citep{ten2012review}. Here we follow the similar set of assumptions in \cite{imai2010identification} and \cite{forastiere2018principal}.

	%
	%
	The first assumption extends the standard ignorability assumption and rules out the unmeasured treatment-outcome confounding.  
	\begin{assumption}[Ignorability]
		\label{A.1}
		Conditional on the observed covariates, the treatment is unconfounded with respect to the potential mediator process and the potential outcomes process:
		\begin{eqnarray*}
			\{\mathbf{Y}_{i}^{t}(1,\mathbf{m}),\mathbf{Y}_{i}^{t}(0,\mathbf{m}),\mathbf{M}_{i}^{t}(1),\mathbf{M}_{i}^{t}(0) \}\independent Z_{i}\mid \mathbf{X}_{i}^{t},
		\end{eqnarray*}
		for any $t$ and $\mathbf{m}\in \mathcal{R}^{[0,t]}$.
	\end{assumption}
	In our context, Assumption \ref{A.1} indicates that there is no unmeasured confounding, besides the observed covariates, between the sources of early adversity and the processes of social bonds and GCs. In other words, early adversity is randomized among the baboons with the same covariates. This assumption is plausible given the early adversity events considered in this study are largely imposed by nature.
	
	The second assumption extending the sequential ignorability assumption in \cite{imai2010identification,forastiere2018principal} to the functional data setting.
	\begin{assumption}[Sequential Ignorability]
		\label{A.2}There exists $\varepsilon>0$, such that for any $0<\Delta<\varepsilon$,the increment of the mediator process is independent of the increment of potential outcomes process from time $t$ to $t+\Delta$, conditional on the observed treatment status, covariates and the mediator process up to time $t$:
		\begin{eqnarray*}
			\{Y_{i}^{t+\Delta}(z,\mathbf{m})-Y_{i}^{t}(z,\mathbf{m})\} \independent \{M_{i}^{t+\Delta}(z')-M_{i}^{t}(z')\}\mid \{Z_{i},\mathbf{X}_{i}^{t},\mathbf{M}_{i}^{t}\},
		\end{eqnarray*}
		for any $z,z',0<\Delta<\varepsilon,t,t+\Delta\in [0,T],\mathbf{m}\in \mathcal{R}^{[0,T]}$.
	\end{assumption}
	In our application, Assumption \ref{A.2} implies that conditioning on the early adversity status, covariates, and the potential social bond history up to a given time point, any change in the social bond values within a sufficiently small time interval $\Delta$ is randomized with respect to the change in the potential outcomes. 
	Namely, there are no unobserved mediator-outcomes confounders in a sufficiently small time interval. Though it differs in the specific form, Assumption \ref{A.2} is in essence the same  sequential ignorability assumption used for the regularly spaced observations in \cite{bind2015longitudinals} and \cite{vanderweele2017timevarying}. This is a crucial assumption in mediation analysis, but is strong and generally untestable in practice because it is usually impossible to manipulate the mediator values, even in randomized trials. 
	
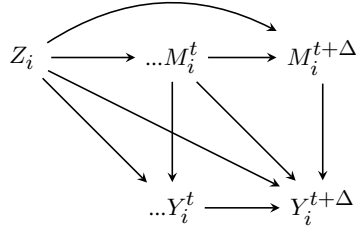
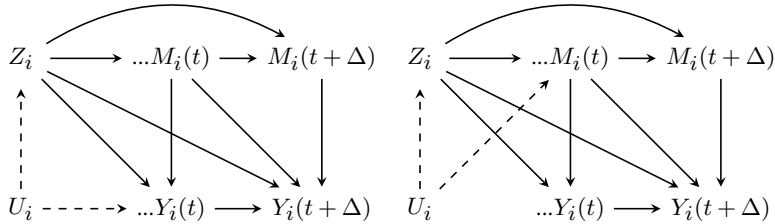
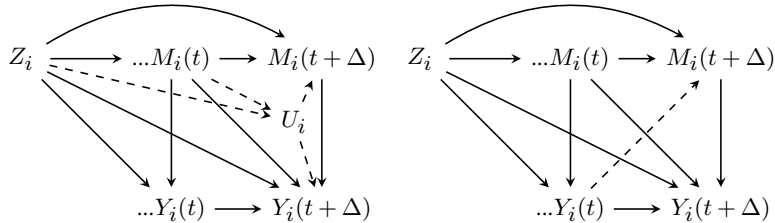
\begin{figure}[!ht]
		\centering
		\begin{subfigure}[b]{\textwidth}
			\centering
			\begin{tikzpicture}
			[
			> = stealth, 
			shorten > = 0.5pt, 
			auto,
			node distance = 2cm, 
			semithick 
			]
			\tikzstyle{every state}=[
			draw = white,
			thick,
			fill = white,
			minimum size = 3mm,
			]
			
			\node[state] (A)  (A){$Z_{i}$};
			\node[right of= A] (B){...$M_{i}^{t}$};
			\node[right of= B] (C) {$M_{i}^{t+\Delta}$};%
			\node[below of= B] (D) {...$Y_{i}^{t}$};%
			\node[below of= C] (E) {$Y_{i}^{t+\Delta}$};%
			
			\path[->] (A) edge node {} (B);
			\path[->] (B) edge node {} (C);
			\path[->] (B) edge node {} (D);
			\path[->] (D) edge node {} (E);
			\path[->] (C) edge node {} (E);
			\path[->] (B) edge node {} (E);
			\path[->] (A) edge node {} (D);
			\path[->] (A) edge node {} (E);
			\path[->] (A) edge  [bend left]  node {} (C);
			\end{tikzpicture}
			\caption{DAG of Assumption \ref{A.1} and \ref{A.2}}
			\label{fig:DAG_A1A2}
		\end{subfigure}
		
		\begin{subfigure}[b]{\textwidth}
			\centering
			\begin{tikzpicture}
			[
			> = stealth, 
			shorten > = 0.5pt, 
			auto,
			node distance = 2cm, 
			semithick 
			]
			\tikzstyle{every state}=[
			draw = white,
			thick,
			fill = white,
			minimum size = 3mm,
			]
			
			\node[state] (A)  (A){$Z_{i}$};
			\node[right of= A] (B){...$M_{i}(t)$};
			\node[right of= B] (C) {$M_{i}(t+\Delta)$};%
			\node[below of= B] (D) {...$Y_{i}(t)$};%
			\node[below of= C] (E) {$Y_{i}(t+\Delta)$};%
			\node[below of= A] (U) {$U_{i}$};%
			
			\path[->] (A) edge node {} (B);
			\path[->] (B) edge node {} (C);
			\path[->] (B) edge node {} (D);
			\path[->] (D) edge node {} (E);
			\path[->] (C) edge node {} (E);
			\path[->] (B) edge node {} (E);
			\path[->] (A) edge node {} (D);
			\path[->] (A) edge node {} (E);
			\path[->] (A) edge  [bend left]  node {} (C);
			\path[dashed,->] (U) edge node {} (A);
			\path[dashed,->] (U) edge node {} (D);
			\end{tikzpicture}
			\begin{tikzpicture}
			[
			> = stealth, 
			shorten > = 0.5pt, 
			auto,
			node distance = 2cm, 
			semithick 
			]
			\tikzstyle{every state}=[
			draw = white,
			thick,
			fill = white,
			minimum size = 3mm,
			]
			
			\node[state] (A)  (A){$Z_{i}$};
			\node[right of= A] (B){...$M_{i}(t)$};
			\node[right of= B] (C) {$M_{i}(t+\Delta)$};%
			\node[below of= B] (D) {...$Y_{i}(t)$};%
			\node[below of= C] (E) {$Y_{i}(t+\Delta)$};%
			\node[below of= A] (U) {$U_{i}$};%
			
			\path[->] (A) edge node {} (B);
			\path[->] (B) edge node {} (C);
			\path[->] (B) edge node {} (D);
			\path[->] (D) edge node {} (E);
			\path[->] (C) edge node {} (E);
			\path[->] (B) edge node {} (E);
			\path[->] (A) edge node {} (D);
			\path[->] (A) edge node {} (E);
			\path[->] (A) edge  [bend left]  node {} (C);
			\path[dashed,->] (U) edge node {} (A);
			\path[dashed,->] (U) edge node {} (B);
			\end{tikzpicture}
			\caption{DAG of two examples of violation to Assumption \ref{A.1} (ignorability)}
			\label{fig:DAG_A1_violation}
		\end{subfigure}
		
		\begin{subfigure}[b]{\textwidth}
			\centering
			\begin{tikzpicture}
			[
			> = stealth, 
			shorten > = 0.5pt, 
			auto,
			node distance = 2cm, 
			semithick 
			]
			\tikzstyle{every state}=[
			draw = white,
			thick,
			fill = white,
			minimum size = 3mm,
			]
			
			\node[state] (A)  (A){$Z_{i}$};
			\node[right of= A] (B){...$M_{i}(t)$};
			\node[right of= B] (C) {$M_{i}(t+\Delta)$};%
			\node[below of= B] (D) {...$Y_{i}(t)$};%
			\node[below of= C] (E) {$Y_{i}(t+\Delta)$};%
			\node[below right = 0.3cm and 0.7cm of  B] (L) {$U_{i}$};%
			
			\path[->] (A) edge node {} (B);
			\path[->] (B) edge node {} (C);
			\path[->] (B) edge node {} (D);
			\path[->] (D) edge node {} (E);
			\path[->] (C) edge node {} (E);
			\path[->] (B) edge node {} (E);
			\path[->] (A) edge node {} (D);
			\path[->] (A) edge node {} (E);
			\path[->] (A) edge  [bend left]  node {} (C);
			\path[dashed,->] (A) edge node {} (L);
			\path[dashed,->] (B) edge node {} (L);
			\path[dashed,->] (L) edge node {} (C);
			\path[dashed,->] (L) edge node {} (E);
			\end{tikzpicture}
			\begin{tikzpicture}
			[
			> = stealth, 
			shorten > = 0.5pt, 
			auto,
			node distance = 2cm, 
			semithick 
			]
			\tikzstyle{every state}=[
			draw = white,
			thick,
			fill = white,
			minimum size = 3mm,
			]
			
			\node[state] (A)  (A){$Z_{i}$};
			\node[right of= A] (B){...$M_{i}(t)$};
			\node[right of= B] (C) {$M_{i}(t+\Delta)$};%
			\node[below of= B] (D) {...$Y_{i}(t)$};%
			\node[below of= C] (E) {$Y_{i}(t+\Delta)$};%
			
			\path[->] (A) edge node {} (B);
			\path[->] (B) edge node {} (C);
			\path[->] (B) edge node {} (D);
			\path[->] (D) edge node {} (E);
			\path[->] (C) edge node {} (E);
			\path[->] (B) edge node {} (E);
			\path[->] (A) edge node {} (D);
			\path[->] (A) edge node {} (E);
			\path[->] (A) edge  [bend left]  node {} (C);
			\path[dashed,->] (D) edge node {} (C);
			\end{tikzpicture}
			\caption{DAG of two examples of violation to Assumption \ref{A.2} (sequential ignorability)}
			\label{fig:DAG_A2_violation}
		\end{subfigure}
		\caption{Directed acyclic graphs (DAG) of Assumptions\ref{A.1},\ref{A.2} and examples of possible violations. The arrows between variables represent a causal relationship, with solid and dashed lines representing measured and unmeasured relationships, respectively. \label{Violation_DAG}}
	\end{figure}

	Assumptions \ref{A.1} and \ref{A.2} are illustrated by the directed acyclic graphs (DAG) in Figure \ref{fig:DAG_A1A2}, which condition on the covariates $\mathbf{X}_{i}^{t}$ and a window between two sufficiently close time points $t$ and $t+\Delta$. The arrows between $Z_{i}$, $M_{i}^{t}$, $Y_{i}^{t}$ represent a causal relationship (i.e., nonparametric structural equation model), with solid and dashed lines representing measured and unmeasured relationships, respectively. Figure \ref{fig:DAG_A1_violation} and \ref{fig:DAG_A2_violation} depicts two possible scenarios where Assumptions \ref{A.1} and \ref{A.2} are violated, respectively, where $U_i$ represents an unmeasured confounder.

	Assumptions \ref{A.1} and \ref{A.2} allow nonparametric  identification of the TE and ACME from the observed data, as summarized in the following theorem.  
	\begin{theorem}
		\label{T.1}
		Under Assumption \ref{A.1},\ref{A.2}, and some regularity conditions (specified in the Supplement Material \citep{aoas_supplement_A}), the TE, ACME and ANDE can be identified nonparametrically from the observed data: for $z=0,1$, we have
		\begin{eqnarray*}
			\tau_{\TE}^{t}&=&\int_{X}  \{E(Y_{i}^{t}|Z_{i}=1,\mathbf{X}_{i}^{t}=\mathbf{x}^{t})-E(Y_{i}^{t}|Z_{i}=0,\mathbf{X}_{i}^{t}=\mathbf{x}^{t})\}\textup{dF}_{\mathbf{X}_{i}^{t}}(\mathbf{x}^{t}),\\
			\tau_{\ACME}^{t}(z)&=&\int_{\mathcal{X}} \int_{R^{[0,t]}} E(Y_{i}^{t}|Z_{i}=z,\mathbf{X}_{i}^{t}=\mathbf{x}^{t}, \mathbf{M}_{i}^{t}=\mathbf{m}) \textup{dF}_{\mathbf{X}_{i}^{t}}(\mathbf{x}^{t})\times\\
			&& \quad\quad \quad  \textup{d}\{\textup{F}_{\mathbf{M}_{i}^{t}|Z_{i}=1,\mathbf{X}_{i}^{t}=\mathbf{x}^{t}}(\mathbf{m})
			-\textup{F}_{\mathbf{M}_{i}^{t}|Z_{i}=0,\mathbf{X}_{i}^{t}=\mathbf{x}^{t}}(\mathbf{m})\},
		\end{eqnarray*}
		where $F_{W}(\cdot)$ and $F_{W|V}(\cdot)$ denotes the cumulative distribution of a random variable or a vector $W$ and the conditional distribution given another random variable or vector $V$, respectively.
	\end{theorem}
	The proof of Theorem \ref{T.1} is provided in the Supplement Material \citep{aoas_supplement_A}. Theorem \ref{T.1} implies that estimating the causal effects requires modeling two components: (a) the conditional expectation of observed outcome process given the treatment, covariates, and the observed mediator process,  $E(Y_{i}^{t}|Z_i,  \mathbf{X}_{i}^{t}, \mathbf{M}_{i}^{t})$, and (b) the distribution of the observed mediator process given the treatment and the covariates, $\textup{F}_{\mathbf{M}_{i}^{t}|Z_i, \mathbf{X}_{i}^{t}}(\cdot)$. These two components correspond to the two linear structural equations in the classic mediation framework of  \cite{baron1986moderator}. In the setting of functional data,  we can employ more flexible models instead of linear regression models, and express the TE and ACME as functions of the model parameters. Theorem \ref{T.1} can be readily extended to more general scenarios such as discrete (i.e., as opposed to continuous) mediators and time-to-event outcomes.

	\section{Modeling mediator and outcome via functional principal component analysis} \label{Sec_Modeling}
	In this section, we propose to employ the functional principal component analysis (FPCA) approach to infer the mediator and outcome processes from sparse and irregular observations \citep{yao2005functional,jiang2010covariatefpca,jiang2011functional}. In order to take into account the uncertainty due to estimating the functional principal components \citep{goldsmith2013corrected},  we adopt a Bayesian model to jointly estimate the principal components and the structural equation models. Specifically, we impose a Bayesian FPCA model similar to that in  \cite{kowal2020bayesian} to project the observed mediator and outcome processes into lower-dimensional representations and then take the first few dominant principal components as the predictors in the structural equation models.   
	
	We assume the potential processes for mediators $\mathbf{M}_{i}^{t}(z)$ and outcomes $\mathbf{Y}_{i}^{t}(z,\mathbf{m})$ have the following Karhunen-Loeve decomposition,
	\begin{gather}
		\label{Mediator_Process}
		M_{i}^{t}(z)=\mu_{M}(\mathbf{X}_{i}^{t})+\sum_{r=1}^{\infty}\zeta_{i,z}^{r}\psi_{r}(t),\\
		\label{Outcome_Process}
		Y_{i}^{t}(z,\mathbf{m})=\mu_{Y}(\mathbf{X}_{i}^{t})+\int_{0}^{t}\gamma(s,t)\mathbf{m}(s)ds+\sum_{s=1}^{\infty}\theta_{i,z}^{s}\eta_{s}(t).
	\end{gather}
	where $\mu_{M}(\cdot)$ and $\mu_{Y}(\cdot)$ are the mean functions of the mediator process $\mathbf{M}_{i}^{t}$ and outcome process $\mathbf{Y}_{i}^{t}$, respectively; $\mathbf{\psi}_{r}(t)$ and $\mathbf{\eta}_{s}(t)$ are the Normal orthogonal eigenfunctions for $\mathbf{M}_{i}^{t}$ and $\mathbf{Y}_{i}^{t}$, respectively, and $\zeta_{i,z}^{r}$ and $\theta_{i,z}^{s}$ are the corresponding principal scores of unit $i$. The above model assumes that the treatment affects the mediation and the outcome processes only through the principal scores. We represent the mediator and outcome process of each unit with its principal score $\zeta_{i,z}^{r}$ and $\theta_{i,z}^{s}$. Given the principal scores , we can transform back to the smooth process with a linear combination. As such, if we are interested in the differences in the process, it is equivalent to investigate the difference in the principal scores. Also, as we usually require only 3 or 4 components to explain most of the variation, we reduce the dimensions of the trajectories effectively by projecting the difference to the principal scores. With the model specification in \eqref{Outcome_Process}, we make an implicit assumption that the ACME and ANDE are the same in the treatment and control groups in our application, $\tau_{\ACME}^{t}(0)=\tau_{\ACME}^{t}(1),\tau_{\ANDE}^{t}(0)=\tau_{\ANDE}^{t}(1)$, and thus there are no interactions between the treatment and the mediator. This assumption leads to a unique decomposition of the TE for simple interpretations \citep{vanderweele2014unification}. 
	
	The underlying processes $\mathbf{M}_{i}^{t}$ and $\mathbf{Y}_{i}^{t}$ are not directly observed. Instead, we assume the observations $M_{ij}$'s and $Y_{ij}$'s are randomly sampled from the respective underlying processes with errors. For the observed mediator trajectories, we posit the following model that truncates to the first $R$ principal components of the mediator process: 
	\begin{gather}
		\label{eq:mediator_model}
		M_{ij}=X_{ij}'\beta_{M}+\sum_{r=1}^{R}\zeta_{i}^{r}\psi_{r}(t_{ij})+\varepsilon_{ij},\quad \varepsilon_{ij}\sim \mathcal{N}(0,\sigma_{m}^{2}),
	\end{gather}
	where $\psi_{r}(t)$ ($r=1,..., R$) are the orthogonormal principal components, $\zeta_{i}^{r}$ ($r=1,..., R$) are the corresponding principal scores, and $\varepsilon_{ij}$ is the measurement error. With similar parametrization that used in \cite{kowal2020bayesian}, we express the principal components as a linear combination of the spline basis $\mathbf{b}(t)=(1,t,b_{1}(t),\cdots,b_{L}(t))'$ in $L+2$ dimensions and choose the coefficients $\mathbf{p}_{r}\in\mathcal{R}^{L+2}$ to meet the normal orthogonality constraints of the $r$th principal component: 
	\begin{gather}
		\label{eq:mediator_pc}
		\psi_{r}(t)=\mathbf{b}(t)'\mathbf{p}_{r}, \text{ subject to} \int_{0}^{T}\psi_{r}^{2}(t)dt=1,\int_{0}^{T}\psi_{r'}(t)\psi_{r''}(t)dt=0, r'\neq r''.
	\end{gather}
	We assume the principal scores $\zeta_{i}^{r}$ are randomly drawn from normal distributions with different means in the treatment and control groups, $\chi_{1}^{r}$ and $\chi_{0}^{r}$, and diminishing variance as $r$ increases: 
	\begin{gather}	
		\label{eq:mediator_pc_distribution}
		\zeta_{i}^{r}\sim \mathcal{N}(\chi_{Z_{i}}^{r}, \lambda_{r}^{2}), \quad \lambda_{1}^{2}\geq\lambda_{2}^{2}\geq\cdots\lambda_{R}^{2}\geq 0.
	\end{gather}
	We select the truncation term $R$ based on the fraction of explained variance (FEV), $\sum_{r=1}^{R}\lambda_{r}^{2}/\sum_{r=1}^{\infty}\lambda_{r}^{2}$ being greater than $90\%$. 
	
	For the observed outcome trajectories, we posit a similar model that truncates to the first $S$ principal components of the outcome process: 
	\begin{gather}
		\label{eq:outcome_model}
		Y_{ij}=X_{ij}^{T}\beta_{Y}+\int_{0}^{t_{ij}} \gamma(u,t) M_{i}^{u}\textup{d}u+\sum_{s=1}^{S}\eta_{s}(t)\theta_{i}^{s}+\nu_{ij}, \quad \nu_{ij}\sim N(0,\sigma_{y}^{2}).
	\end{gather}
	We express the principal components $\eta_{s}$ as a linear combination of the spline basis $\mathbf{b}(t)$, with the normal orthogonality constraints:
	\begin{gather}
		\label{eq:outcome_pc}
		\eta_{s}(t)=\mathbf{b}(t)'\mathbf{q}_{s},  \text{ subject to } \int_{0}^{T}\eta_{s}(t)^{2}dt=1,\int_{0}^{T}\eta_{s'}(t)\eta_{s''}(t)dt=0, s'\neq s''.
	\end{gather}
	Similarly, we assume that the principal scores of the outcome process for each unit come from two different normal distributions in the treatment and control group with  means $\xi_{1}^{s}$ and $\xi_{0}^{s}$ respectively, and a shrinking variance $\rho_{s}^{2}$:
	\begin{gather}
		\label{eq:outcome_pc_distribution}
		\theta_{i}^{s}\sim\mathcal{N}(\xi_{Z_{i}}^{s},\rho_{s}^{2}), \quad \rho_{1}^{2}\geq \rho_{2}^{2}\geq \cdots\rho_{S}^{2}\geq 0.
	\end{gather}
	We select the truncation term $S$ based on the FEV being greater than $90\%$, namely $\sum_{s=1}^{S}\rho_{s}^{2}/\sum_{s=1}^{\infty}\rho_{s}^{2}\geq 90\%$.
	
	We assume the effect of the mediation process on the outcome is concurrent, namely the outcome process at time $t$ does not depend on the past value of the mediation process. As such, $\gamma(u,t)$ can be shrunk to $\gamma$ instead of the integral in Model \eqref{eq:outcome_model},
	\begin{gather}
		\label{eq:outcome_model_concurrent}
		Y_{ij}=X_{ij}^{T}\beta_{Y}+\gamma M_{ij}+\sum_{s=1}^{S}\eta_{s}(t)\theta_{i}^{s}+\nu_{ij},\quad \nu_{ij}\sim N(0,\sigma_{y}^{2}).
	\end{gather}
	
	The causal estimands, the TE and ACME, can be expressed as functions of the parameters in the above mediator and outcome models:
	\begin{eqnarray}
		\label{TT_EXPRESS}
		\tau_{\TE}^t&=&\sum_{s=1}^{S}(\xi_{1}^{s}-\xi_{0}^{s})\eta_{s}(t)+\gamma\sum_{r=1}^{R} (\chi_{1}^{r}-\chi_{0}^{r})\psi_{r}(t),\\
		\label{IT_EXPRESS}
		\tau_{\ACME}^{t}&=&\gamma \sum_{r=1}^{R}(\chi_{1}^{r}-\chi_{0}^{r})\psi_{r}(t).
	\end{eqnarray}
	
	To account for the uncertainty in estimating the above models, we adopt the Bayesian paradigm and impose prior distributions for the parameters \citep{kowal2020bayesian}. For the basis function $\mathbf{b}(t)$ to construct principal components, we choose the thin-plate spline which takes the form $\mathbf{b}(t)=(1,t,(|t-k_{1}|)^{3},\cdots,|t-k_{L}|^{3})'\in \mathcal{R}^{L+2}$, where the $k_{l} \ (l=1,2,\cdots,L)$ are the pre-defined knots on the time span. We set the values of knots $k_{l}$ with the quantiles of observation time grids. For the parameters of the principal components, taking the mediator model as an example, we impose the following priors on the parameters in \eqref{eq:mediator_pc}:
	\begin{gather*}
		\mathbf{p}_{r}\sim N(0,h_{r}^{-1}\Omega^{-1}),h_{r}\sim \textup{Uniform}(\lambda^{2}_{r},10^{4}),
	\end{gather*}
	where $\Omega\in \mathcal{R}^{(L+2)\times(L+2)}$ is the roughness penalty matrix and $h_{r}>0$ is the smooth parameter. The implies a Gaussian Process prior on $\psi_{r}(t)$ with mean function zero and covariance function $\textup{Cov}(\psi_{r}(t),\psi_{r}(s))=h_{r}\mathbf{b}'(s)\Omega\mathbf{b}(t)$. We choose the $\Omega$ such that $[\Omega_{r}]_{l,l'}=(k_{l}-k_{l})^{2}$,when $l,l'>2,$ and $[\Omega_{r}]_{l,l'}=0$ when $l,l'\leq 2$. For the distribution of principal scores in \eqref{eq:mediator_pc_distribution}, we specify a multiplicative Gamma prior \citep{bhattacharya2011sparse,montagna2012bayesian} on the variance to encourage shrinkage as $r$ increases,
	\begin{gather*}
		\chi_{0}^{r},\chi_{1}^{r}\sim N(0,\sigma^{2}_{\chi_{r}}),\quad \sigma^{-2}_{\chi_{r}}=\prod_{l\leq r}\delta_{\chi_{l}},  \delta_{\chi_{1}}\sim \textup{Ga}(a_{\chi_{1}},1 ), \delta_{\chi_{l}}\sim \textup{Ga}(a_{\chi_{2}},1),l\geq 2,\\
		\lambda_{r}^{-2}=\prod _{l\leq r}\delta_{l},\quad \delta_{1}\sim \textup{Ga}(a_{1},1), \quad \delta_{l}\sim \textup{Ga}(a_{2},1), l\geq 2,\\
		a_{1},a_{\chi_{1}}\sim \textup{Ga}(2,1), \quad a_{2},a_{\chi_{2}}\sim \textup{Ga}(3,1).
	\end{gather*}
	Further details on the hyperparameters of the priors can be found in \cite{bhattacharya2011sparse} and \cite{durante2017note}. For the coefficients of covariates $\beta_{M}$, we specify a diffused normal prior $\beta_{M}\sim \mathcal{N}(0,100^{2}*\textup{I}_{\textup{dim}(X)})$. We impose similar prior distributions for the parameters in the outcome model.
	
	Posterior inference can be obtained by Gibbs sampling. The credible intervals of the causal effects $\tau_{\TE}^t$ and $\tau_{\ACME}^{t}$ can be easily constructed using the posterior sample of the parameters in the model. Details of the Gibbs sampler are provided in the Supplement Material \cite{aoas_supplement_A}.

	\section{Empirical Application}\label{Sec_Application}
	\subsection{Results of FPCA}
\begin{figure}[ht]
		\centering
		\includegraphics[width=0.4\textwidth]{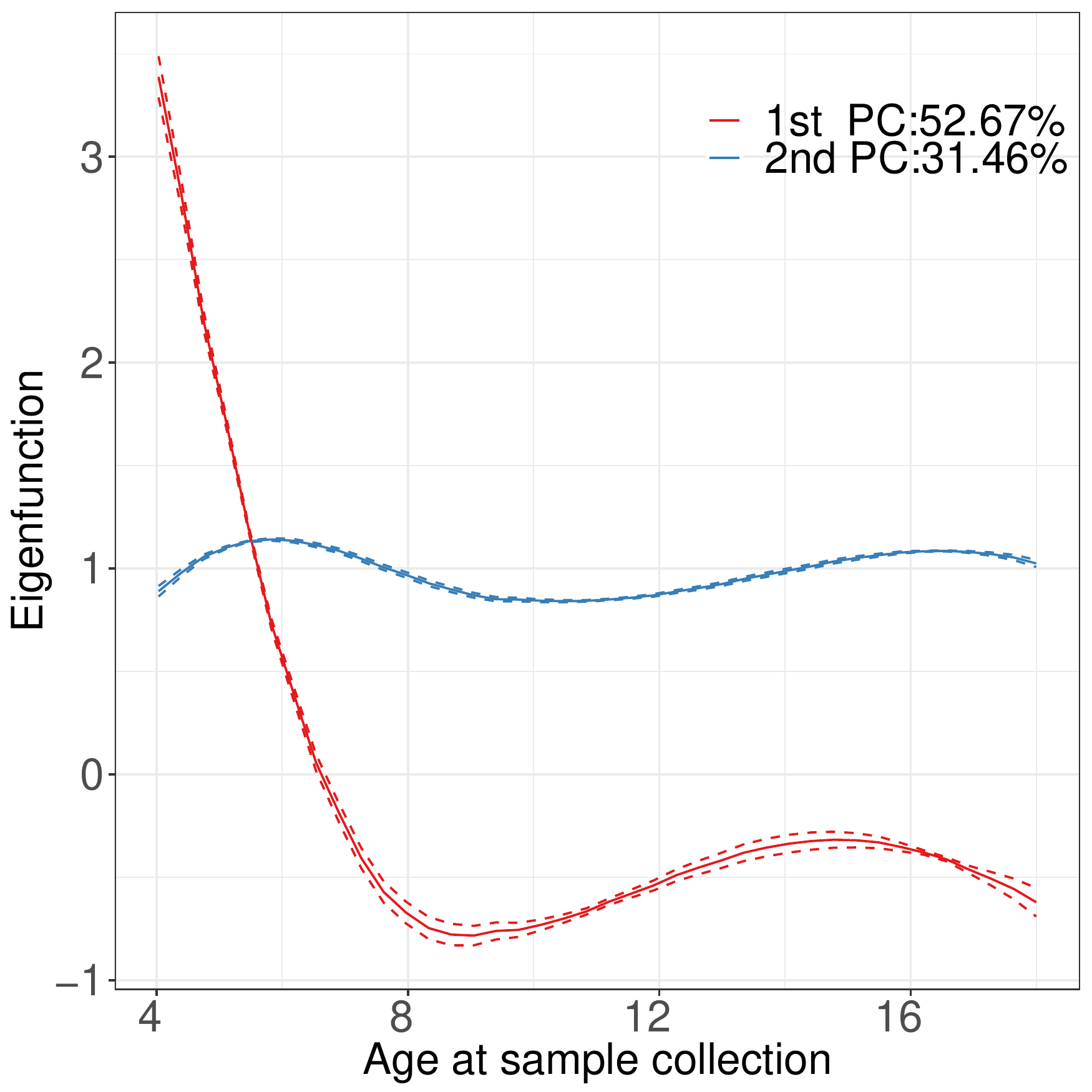}
		\hspace{2em}
		\includegraphics[width=0.4\textwidth]{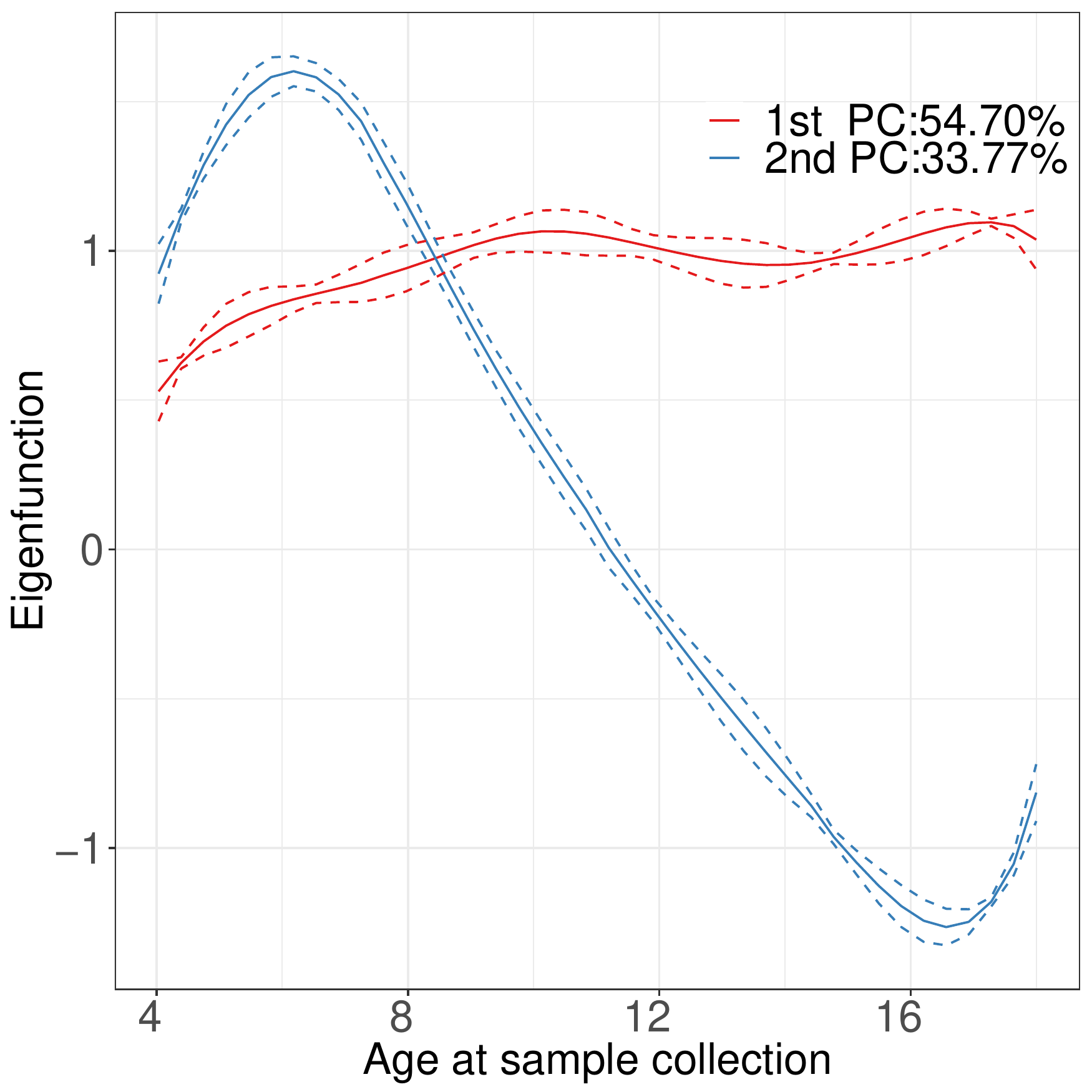}
		\caption{The first two functional principal components of the process of the mediator, i.e. social bonds (left panel) and the outcome, i.e., GC concentrations (right panel).}
		\label{fig:pc}
	\end{figure}

	We apply the method and models proposed in Section \ref{Sec_Framework} and \ref{Sec_Modeling} to the data described in Section \ref{sec:data_general} to investigate the causal relationship between early adversity, social bonds and stress in wild baboons. Here we first summarize the results of FPCA of the observed trajectories. We posit model \eqref{eq:mediator_model} for the social bonds and Model \eqref{eq:outcome_model_concurrent} for the GC concentrations, with some modifications. First, we added two random effects, one for social group and one for hydrological year, in both models. Second, in the outcome model, we use the log transformed GC concentrations instead of the original scale as the outcome, which allows us to interpret the coefficient as the percent difference in GC concentrations between the treatment and control groups. 
	For both the mediator and outcome processes, the first three functional principal components explain more than 90\% of the total variation, and thus we use them in the structural equation model for mediation analysis. Figure \ref{fig:pc} shows the first two principal components extracted from the mediator (left panel) and outcome (right panel) processes. For the social bond process, the first two principal components explain 53\% and 31\% of the total variation, respectively. The first component depicts a drastic change in the early stage of a baboon's life and stabilizes afterwards. The second component is relatively stable across the life span. For the GC process, the first two functional principal components explain 54\% and 34\% of the total variation, respectively. The first component depicts a stable trend throughout the life span. The second component shows a quick rise, then steady drop pattern across the lifespan.
	
\begin{figure}[ht]
		\centering
		\includegraphics[width=0.4\textwidth]{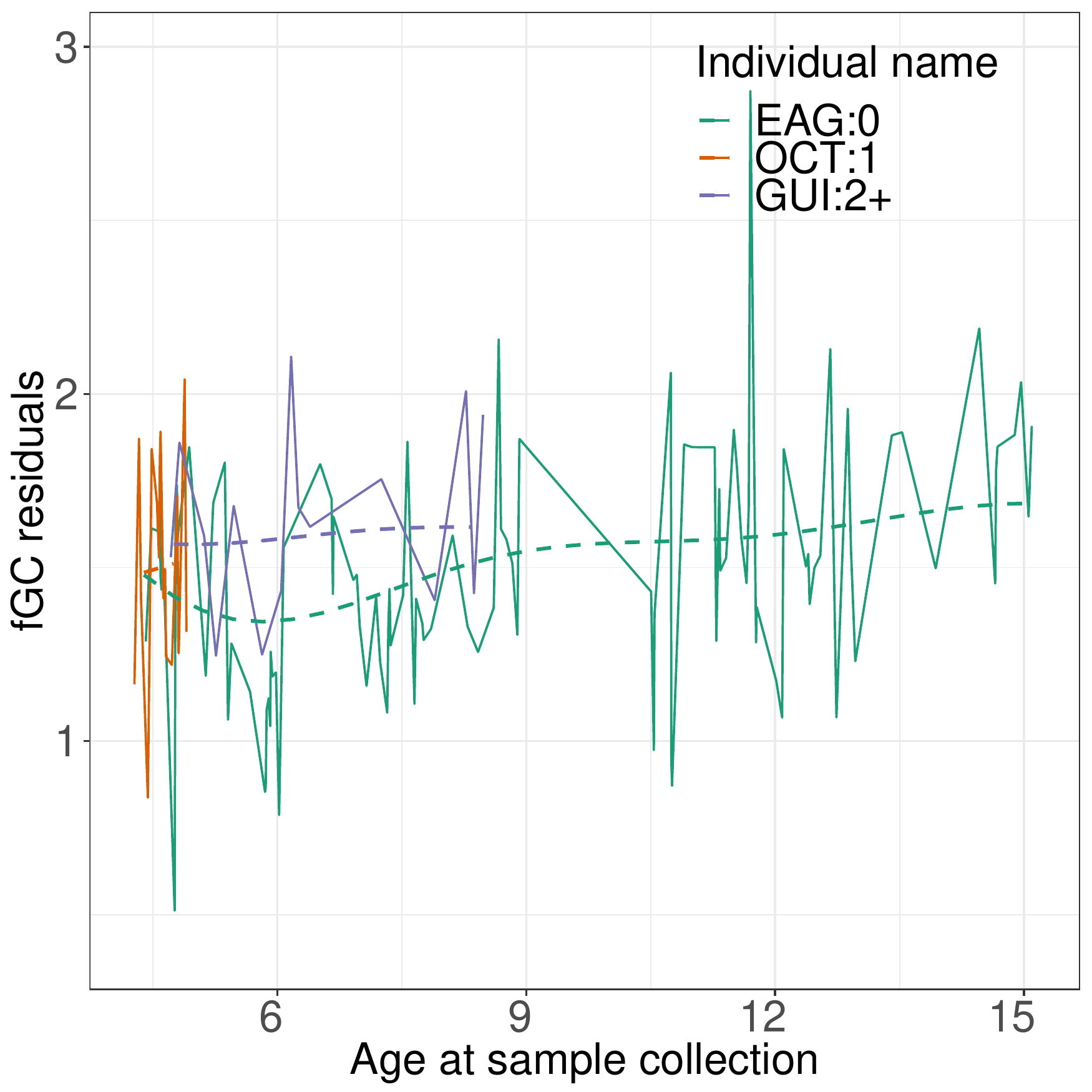}
		\hspace{2em}
		\includegraphics[width=0.4\textwidth]{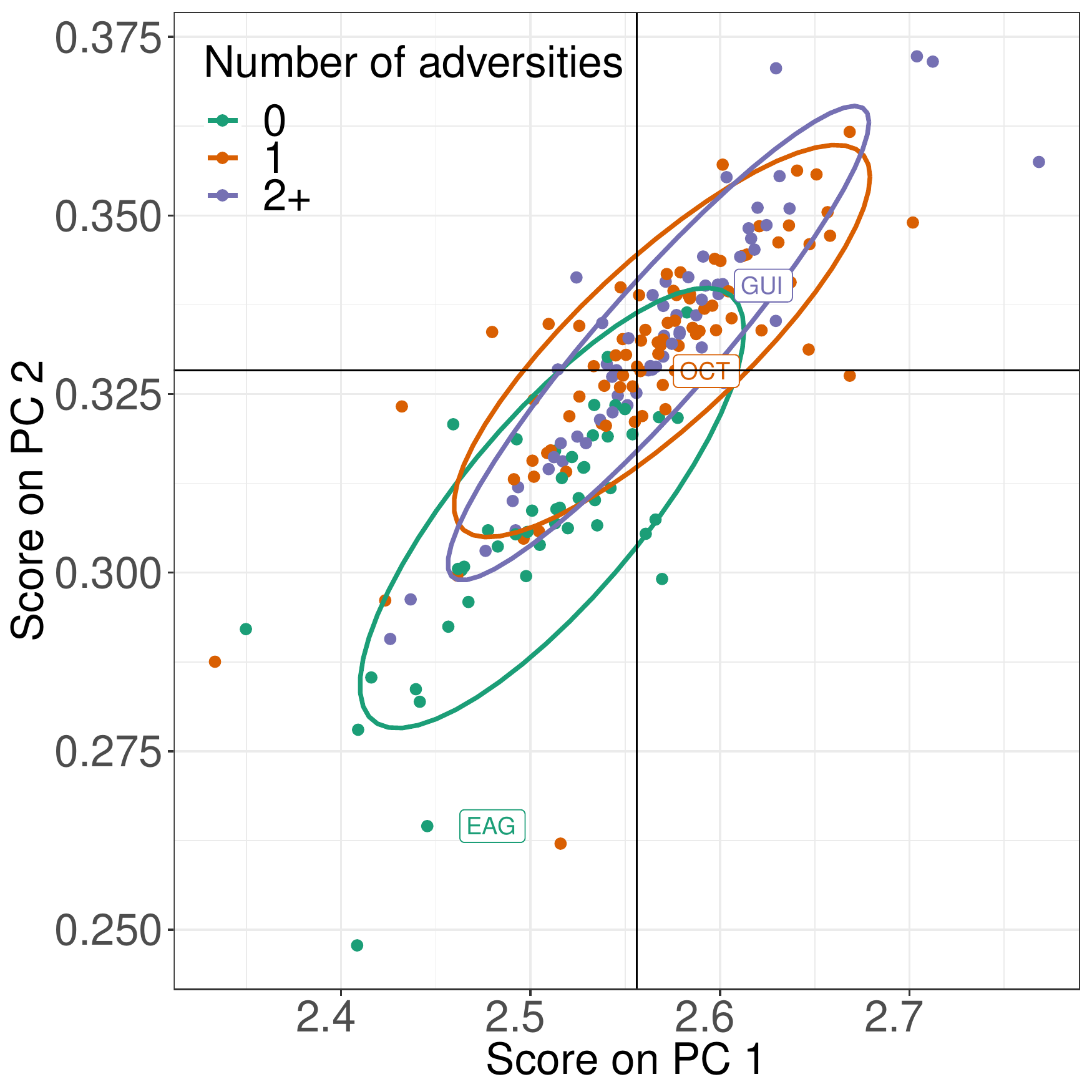}
		\caption{Left panel: Observed trajectory of GCs versus the posterior mean of its imputed smooth process of three baboons who experienced zero (EAG), one (OCT) and two (GUI) sources of early adversity, respectively. Right panel: Principal scores of the first (X-axis) versus second (Y-axis) principal component for the GC process of all subjects in the sample; plotted in clusters based on the number of early adversities experienced. }
		\label{fig:ordination}
	\end{figure}

	The left panel of Figure \ref{fig:ordination} displays the observed trajectory of GCs versus the posterior mean of the imputed smooth process of three baboons who experienced zero (EAG), one (OCT), and two (GUI) sources of early adversity, respectively. We can see that the imputed smooth process generally captures the overall time trend of each subject while reducing the noise in the observations. The pattern is similar for the animals' social bonds, which is shown in the Supplement Material \citep{aoas_supplement_A} with a few more randomly selected subjects. Recall that each subject's observed trajectory is fully captured by its vector of principal scores, and thus the principal scores of the first few dominant principal components adequately summarize the whole trajectory.  The right panel of Figure \ref{fig:ordination} shows the principal scores of the first (X-axis) versus second (Y-axis) principal component for the GC process of all subjects in the sample, plotted in clusters based on the number of early adversities experienced. We can see that significant differences exist in the distributions of the first two principal scores between the group who experienced no early adversity and the groups experienced one or more sources of adversity.

	\subsection{Results of causal mediation analysis}
	
	
	We perform a separate causal mediation analysis for each source of early adversity. Table \ref{tab:results} presents the posterior mean and 95\% credible interval of
	the total effect (TE), direct effect (ANDE) and indirect effect  mediated through social bonds (ACME) of each source of early adversity on adult GC concentrations, as well as the effects of early adversity on the mediator (social bonds). First, from the first column of Table \ref{tab:results} we can see that experiencing any source of early adversity would reduce the strength of a baboon's social bond strength with other baboons in adulthood. The negative effect is particularly severe for those who experienced drought, high group density, or maternal death in early life. For example, compared with the baboons who did not experience any early adversity, the baboons who experienced maternal death have a $0.221$ unit decrease in social bonds, translating to a $0.4$ standard deviation difference in social bond strength in this population. Overall, experiencing at least one source of early adversity corresponds to social bonds that are $0.2$ standard deviations weaker in adulthood.
	
	Second, from the second column of Table \ref{tab:results} we can see a strong total effect of early adversity on female baboon's GC concentrations across  adulthood. Baboons who experienced at least one source of adversity had GC concentrations that were approximately 9\% higher than their peers who did not experience any adversity. 
	Although the range of total effect sizes across all individual adversity sources varies from 4\% to 14\%, the point estimates are consistently toward higher GC concentrations, even for the early adversity sources for which the credible interval includes zero. Among the individual sources of adversity, females who were born during a drought, into a high-density group, or to a low-ranking mother had particularly elevated GC concentrations (12-14\%) in adulthood, although the credible interval of high group density includes zero.  
	\begin{table}[!ht]
		\centering
		\caption{Total, direct and indirect causal effects of individual and cumulative sources of early adversity on social bonds and GC concentrations in adulthood in wild female baboons. 95\% credible intervals are in the parenthesis.}
		\resizebox{\textwidth}{!}{
			\begin{tabular}{lccccc}
				\hline
				Source of adversity &effect on mediator & $\tau_{\TE}$ & $\tau_{\ACME}$ &$\tau_{\ANDE}$ \\
				\hline
				Drought  &$-0.164$&$0.124$&$0.009$&$0.114$\\
				&$(-0.314,-0.014)$&$(0.007,0.241)$&	$(0.000,0.017)$ & $(0.005,0.222)$\\
				Competing sibling&$-0.106$&$0.084$&	$0.006$ & $0.078$
				\\
				&	$(-0.249,0.030)$&$(-0.008,0.172)$&$(0.003,0.009)$&
				$(-0.012,0.163)$\\
				High group density	&$-0.271$	&$0.123$&	$0.015$ & $0.108$\\
				&$(-0.519,-0.023)$&$(-0.052,0.281)$& 	$(0.000,0.029)$& $(-0.053,0.252)$
				\\
				Maternal death&$-0.221$&$0.061$	&$0.011$& $0.049$
				\\
				&$(-0.423,-0.019)$&$(-0.006,0.129)$&		$(0.005,0.014)$&$(	-0.014,0.113)$\\
				Low maternal rank&$-0.052$&$0.134$	&	$0.008$ & $0.126$
				\\
				&$(-0.298,0.001)$&$(0.011,0.256)$&	$(0.005,0.011)$&$(0.008,0.244)$
				\\
				Maternal social isolation&	$-0.040$&$0.035$&	$0.002$&$0.033$\\
				&	$(-0.159,0.095)$&$(-0.045,0.116)$&	$(0.000,0.005)$&$(-0.044,0.111)$
				\\
				At least one&$-0.102$&$0.092$&$0.007$&$0.084$\\
				&	$(-0.195,-0.008)$	&$(0.005,0.178)$&	$(0.002,0.009)$& $(0.009,0.159)$
				\\
				\hline
			\end{tabular}
		}
		\label{tab:results}
	\end{table}

	Third, while female baboons who experienced harsh conditions in early life show higher GC concentrations in adulthood, we found no evidence that these effects were significantly mediated by the absence of strong social bonds. Specifically, the mediation effect $\tau_{\ACME}$ (third column in Table \ref{tab:results}) is consistently small; the strength of females’ social bonds with other females accounted for a difference in GCs of only 0.85\% when averaged across the six individual adversity sources, even though the credible intervals did not include zero for five of the six individual adversity sources. On the other hand, the direct effects $\tau_{\ANDE}$ (fourth column in Table \ref{tab:results}) are much stronger than the mediation effects. When averaged across the six adversity sources, the direct effect of early adversity on GC concentrations was 11.6 times stronger than the mediation effect running through social bonds. For example, for females who experienced at least one source of early adversity, the direct effect explain an 8.4\% difference in GC concentrations, while the mediation effect only takes up 0.7\% for the difference in GCs.  
	
	We also assess the plausibility of the key causal assumptions in the application. One possible violation can be due to `feedback' between the social bond and GC processes, as is shown in Figure \ref{fig:DAG_A2_violation}. We performed a sensitivity analysis by adding (a) the most recent prior observed GC value, or (b) the average of all past observed GC values, as a predictor in the mediation model, which led to little difference in the results and thus bolsters sequential ignorability. 
	Though we are not aware of the existence of other sequential confounders, we also cannot rule them out. 
	
	The above findings on the causal relationships among early adversity, social bonds, and GC concentrations in wild baboons are compatible with observations in many other species that early adversity and weak relationships both give rise to poor health, and that early adversity predicts various forms of social dysfunction, including weaker relationships. However, they call into question the notion that social bonds play a major role in mediating the effect of early adversity on poor health. In wild female baboons, any such effect appears to be functionally biologically irrelevant or minor.
	
	\section{Simulations\label{Sec_Simulation}}
	In this section, we conduct simulations to further evaluate the operating characteristics of the proposed method and compare it with two standard methods. 
	\subsection{Simulation design} \label{sec:simu_design}
	We generate 200 units to approximate the sample size in our application. For each unit, we make $T_{i}$ observations at the time grid $\{t_{ij}\in [0,1], j=1,2,\cdots,T_{i}\}$. We draw $T_{i}$ from a Poisson distribution with mean $T$ and randomly pick $t_{ij}$ uniformly:
	\begin{gather*}
		T_{i}\sim \textup{Poisson}(T), \quad t_{ij}\sim \textup{Uniform}(0,1),  \quad j=1,2,\cdots,T_{i}.
	\end{gather*}
	For each unit $i$ and time $j$, we generate three covariates from a tri-variate Normal distribution, $\mathbf{X}_{ij}=(X_{ij1},X_{ij2}, X_{ij3}) \sim \mathcal{N}([0,0,0]^{T},\sigma_{X}^{2}\textup{I}_{3})$. We simulate the binary treatment indicator from $Z_i=\one\{c_{i1}> 0\}$,
	where $c_{i1}\sim \mathcal{N}(0,1)$. To simulate the sparse and irregular mediator trajectories, we first simulate a smooth underlying process $M_{i}^{t}(z)$ for the mediators:
	\begin{gather*}
		M^{t}_{i}(z)=0.2+\{0.2+2t+\textup{sin}(2\pi t)\})(z+1)-X_{ij1}+0.5X_{ij2}+
		\varepsilon_{i}^{m}(t)+c_{i2},
	\end{gather*}
	where the error term $\varepsilon_{i}^{m}(t)\sim \textup{GP}(0,\sigma_{m}^{2}\textup{exp}\{-8(s-t)^{2}\})$ is drawn from a Gaussian Process (GP) with an exponential kernel and $\sigma_{m}^{2}$ controlling the volatility of the realized curves, and $c_{i2}\sim\mathcal{N}(0,\sigma_{m}^{2})$ to represent the individual random intercepts.  The mean value of the mediator process depends on the covariates and time index $t$. The polynomial term and the trigonometric function of $t$ introduce the long term growth trend and periodic fluctuations, respectively. Also, the coefficient of $z$ evolves as the time changes, implying a time-varying treatment effect on the mediator. Similarly, we specify a GP model for the outcome process,
	\begin{eqnarray*}
		Y_{i}^{t}(z,\mathbf{m})&=&\mathbf{m}^{t}+\textup{cos}(2\pi t)+0.1t^{2}+2t+\{\textup{cos}(2\pi t)+0.2t^{2}+3t\}z-\\
		&& \quad 0.5 X_{ij2}+X_{ij3}+
		\varepsilon_{i}^{y}(t)+c_{i3}, 
	\end{eqnarray*}
	where the error term $\varepsilon_{i}^{y}(t)\sim \textup{GP}(0,\sigma_{y}^{2}\textup{exp}\{-8(s-t)^{2}\})$ is drawn from a GP, and $c_{i3}\sim\mathcal{N}(0,\sigma_{y}^{2})$ controls the individual random effects for the outcome process. 
	
	
	The above settings imply non-linear true causal effects ($\tau_{\TE}^t$ and $\tau_{\ACME}^{t}$) in time, which are shown as the dashed lines in Figure \ref{fig:MFPCA}. 
	Upon simulating the processes, we evaluate the potential values of the mediators and outcomes at the sampled time point $t_{ij}$ to obtain the observed trajectories with measurement error: 
	\begin{gather*}
		M_{ij}\sim\mathcal{N}(\mathbf{M}_{i}^{t_{ij}}(Z_{i}),1), \quad
		Y_{ij}\sim\mathcal{N}(\mathbf{Y}_{i}^{t_{ij}}(Z_{i},\mathbf{M}_{i}^{t_{ij}}(Z_{i})),1).
	\end{gather*}
	We control the sparsity of the mediator and outcome trajectories by varying the value of $T$ in the grid of $(15,25,50,100)$, namely the average number of observations for each individual. 
	
	We compare the proposed method in Section \ref{Sec_Modeling} (abbreviated as MFPCA) with two standard methods in longitudinal data analysis: the random effects model \citep{laird1982random} and the generalized estimating equations (GEE) \citep{liang1986longitudinal}. To facilitate the comparisons, we 
	aggregate the time-varying mediation effects into the following scalar values:
	\begin{eqnarray*}
		\tau_{\ACME} = \int_{0}^{T}\tau_{\ACME}^{t}\textup{d}t, \quad \tau_{\TE} = \int_{0}^{T}\tau_{\TE}^{t}\textup{d}t.
	\end{eqnarray*}
	The true values for $\tau_{\ACME}$ and $\tau_{\TE}$ in the simulations are $1.20$ and $2.77$ respectively.
	
	For the random effects approach, we fit the following two models: 
	\begin{eqnarray}
		M_{ij} = X_{ij}^{T}\beta_{M}+s_{m}(T_{ij})+\tau_{m}Z_{i}+r^{m}_{ij}+\varepsilon^{m}_{ij},\\
		Y_{ij} = X_{ij}^{T}\beta_{Y}+s_{y}(T_{ij})+\tau_{y}Z_{i}+\gamma M_{ij}+r^{y}_{ij}+\varepsilon^{y}_{ij},
	\end{eqnarray}
	where $r^{m}_{ij}$ and $r^{y}_{im}$ are normally distributed random effects with zero means, $s_{m}(T_{ij})$ and $s_{y}(T_{ij})$ are thin plate splines to capture the nonlinear effect of time. To model the time dependency, we specify an AR(1) correlation structure for the random effects, thus $\textup{Corr}(r^{m}_{ij},r^{m}_{ij+1})=p_{1},\textup{Corr}(r^{y}_{ij},r^{y}_{ij+1})=p_{2}$, namely the correlation decay exponentially within the observations of a given unit. Given the above random effects model, the mediation effect and TE can be calculated as: $\hat{\tau}_{\ACME}^{\RD}=\hat{\gamma}\hat{\tau}_{m},\hat{\tau}_{\TE}^{\RD}=\hat{\gamma}\hat{\tau}_{m}+\hat{\tau}_{y}$.
	
	For the GEE approach, we specify the following estimation equations:
	\begin{eqnarray}
		E(M_{ij}|X_{ij},Z_{i})=X_{ij}^{T}\beta_{M}+\tau_{m}Z_{i},\\
		E(Y_{ij}|M_{ij},X_{ij},Z_{i})=X_{ij}^{T}\beta_{M}+\tau_{y}Z_{i}+\gamma M_{ij}.
	\end{eqnarray}
	For the working correlation structure, we consider the AR(1) correlation for both the mediators and outcomes.
	Similarly, we obtain the estimations through $\hat{\tau}_{\ACME}^{\GEE}=\hat{\gamma}\hat{\tau}_{m},\hat{\tau}_{\TE}^{\GEE}=\hat{\gamma}\hat{\tau}_{m}+\hat{\tau}_{y}$ with two different correlation structures.
	
	It is worth noting that both the random effects model and the GEE model generally lack the flexibility to accommodate irregularly-spaced longitudinal data, which renders specifying the correlation between consecutive observations difficult. For example, though the AR(1) correlation takes into account the temporal structure of the data, it still requires the correlation between any two consecutive observations to be constant, which is unlikely to be the case in use cases with irregularly-spaced data. Nonetheless, we compare the proposed method with these two models as they are the standard methods in longitudinal data analysis.
	
	\subsection{Simulation results}
	We apply the proposed MFPCA method, the random effects model, and the GEE model in Section \ref{sec:simu_design} to the simulated data $\{Z_{i}, \mathbf{X}_{ij}, M_{ij}, Y_{ij}\}$, to estimate the causal effects $\tau_{\TE}$ and $\tau_{\ACME}$.
	
\begin{figure}[!ht]
		\centering
		\includegraphics[width=0.95\textwidth]{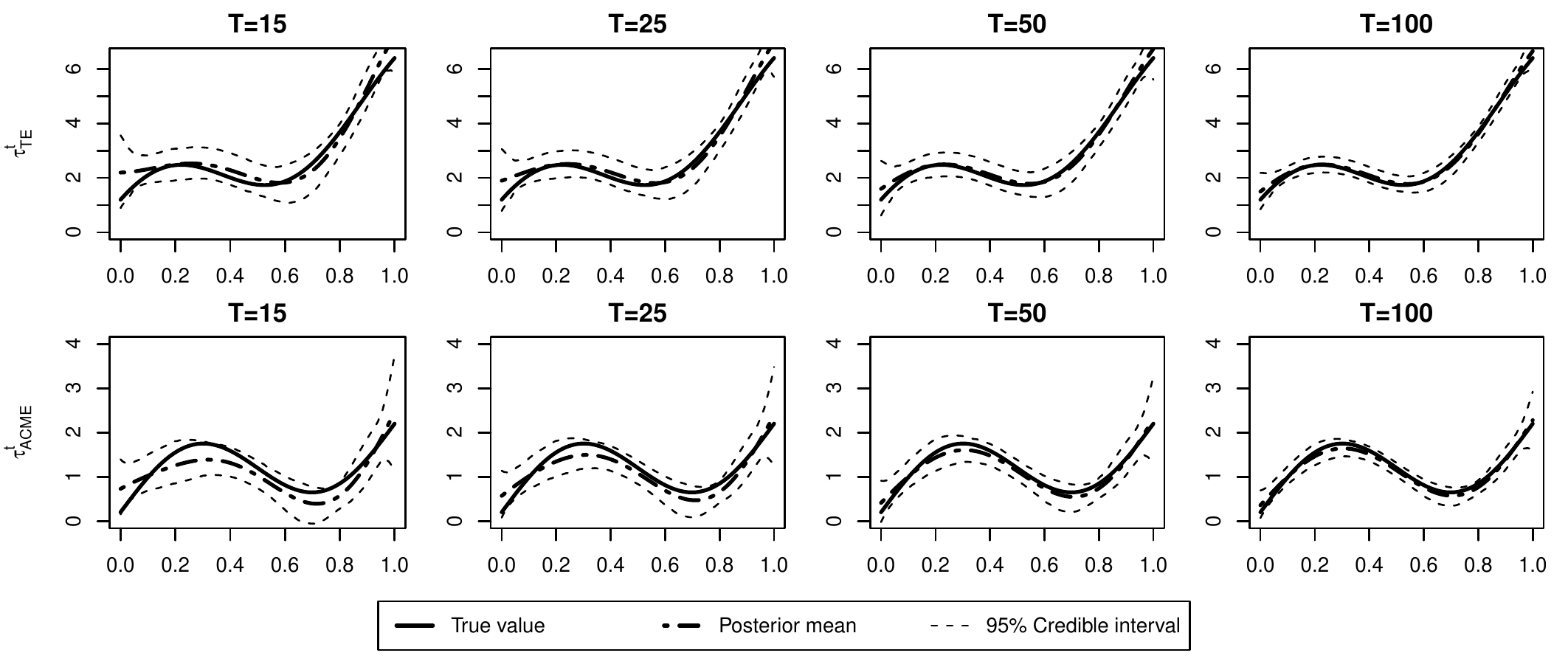}
		\caption{Posterior mean of  $\tau_{\TE}^{t}$,$\tau_{\ACME}^{t}$ and 95\% credible intervals in one simulated dataset under each level of sparsity with $200$ units. The solid lines are the true surfaces for $\tau_{\TE}^{t}$ and $\tau_{\ACME}^{t}$ }
		\label{fig:MFPCA}
	\end{figure}

	Figure \ref{fig:MFPCA} shows the causal effects and associated 95\% credible interval estimated from MFPCA in one randomly selected simulated dataset under each of the four levels of sparsity $T$. Regardless of $T$, MFPCA appears to estimate the time-varying causal effects satisfactorily, with the 95\% credible interval covering the true effects at any time. As expected, the accuracy of the estimation increases as the frequency of the observations increases.   
	
	Table \ref{tab:simu_comparison} presents the absolute bias, root mean squared error (RMSE) and coverage rate of the 95\% confidence interval of $\tau_{\TE}$ and $\tau_{\ACME}$ under the MFPCA, the random effects model and the GEE model based on 1000 simulated datasets for each level of sparsity $T$ in $[15,25,50,100]$. The performance of all three methods improves as the frequency of observations increases. With low frequency ($T<100$), i.e. sparse observations, MFPCA consistently outperforms the random effects model, which in turn outperforms GEE in all measures. The advantage of MFPCA over the other two methods diminishes as the frequency increases. In particular, with dense observations ($T=100$), MFPCA leads to similar results as random effects, though both still outperform GEE. The simulation results bolster the use of our method in the case of sparse data. 
	
	We also conducted the same simulations with larger sample sizes, $N=500, 1000$. MFPCA's advantage over the random effects and GEE models in terms of bias and RMSE increases as the sample size increases. With $N=500$, MFPCA already achieves a coverage rate close to the nominal level. We leave the detailed results to Supplement Material \citep{aoas_supplement_A}.
	
	\begin{table}[!ht]
		\centering
		\caption{Absolute bias, RMSE and coverage rate of the 95\% confidence interval of MFPCA, the random effects model and the generalized estimating equation (GEE) model under different frequency of observations in the simulations. }
		\label{tab:simu_comparison}
		\begin{tabular}{ccccccc}
			\hline
			&\multicolumn{3}{c}{$\tau_{\TE}$} &\multicolumn{3}{c}{$\tau_{\ACME}$}\\
			Method &Bias & RMSE & Coverage&Bias & RMSE & Coverage\\
			\hline
			&\multicolumn{6}{c}{$T$=15}\\
			MFPCA&0.103 &0.154 &88.4\%& 0.134&0.273& 86.4\% \\
			Random effects&0.165&0.208& 78.2\% &0.883&1.673&69.5\%\\
			GEE &0.183&0.304 & 77.6\%&0.987&2.051& 61.8\%\\
			\hline
			&\multicolumn{6}{c}{$T$=25}\\
			MFPCA&0.092 &0.123 &92.3\%& 0.102&0.246& 90.6\% \\
			Random effects&0.124&0.165& 81.2\% &0.679&1.263&72.3\%\\
			GEE &0.152&0.273 & 80.3\%&0.860&1.753& 64.4\%\\
			\hline
			&\multicolumn{6}{c}{$T$=50}\\
			MFPCA&0.087 &0.112 &93.5\%& 0.094&0.195& 92.3\% \\
			Random effects&0.109&0.134& 90.3\% &0.228&0.497&88.8\%\\
			GEE &0.121&0.175 & 83.5\%&0.236&0.493& 80.8\%\\
			\hline
			&\multicolumn{6}{c}{$T$=100}\\
			MFPCA&0.053 &0.089 &94.3\%& 0.064&0.163& 93.1\% \\
			Random effects&0.046&0.093& 93.1\% &0.053&0.154&92.8\%\\
			GEE &0.093&0.124 & 90.5\%&0.098&0.161& 90.3\%\\
			\hline
		\end{tabular}
	\end{table}

	\section{Discussion\label{Sec_Discussion}}
	We proposed a framework for conducting causal mediation analysis with sparse and irregular longitudinal mediator and outcome data. We defined several causal estimands (total, direct and indirect effects) in such settings and specified structural assumptions to nonparametrically identify these effects. For estimation and inference, we combine functional principal component analysis (FPCA) techniques and the standard two structural-equation-model system. In particular, we use a Bayesian FPCA model to reduce the dimensions of the observed trajectories of mediators and outcomes. We applied the proposed method to analyze the causal effects of early adversity on adult social bonds and adult GC hormone concentrations in a sample of wild female baboons. We found that experiencing adversity before maturity generally hampers a baboon's ability to build social bonds with other baboons and increases GC hormone concentrations in adulthood. Chronically elevated stress hormones have been linked to disease in many species. However, the effect of early adversity on adult GC concentrations does not appear to be mediated by the strength of the animals' social bonds, at least in  wild baboons. 
	
	Identification of the causal effects in our method relies a set of structural assumptions. In particular, sequential ignorability plays a key role but it is untestable. In our application, we have adjusted for all important confounders related to the sequential ignorability according to our substantive biological knowledge, but we still cannot rule out the possibility of unobserved confounders. For example, the loss of a close companion would affect a baboon's social bond and stress afterwards, but such data may be not systematically collected. This is a common challenge in causal mediation analysis. Conducting a sensitivity analysis would shed light on the consequences of violating such assumptions \citep{imai2010identification}. However, it is a non-trivial task to design a sensitivity analysis in complex settings such as ours, which usually involves more untestable structural and modeling assumptions. This is probably why sensitivity analysis is rarely performed in the causal mediation analysis literature despite its obvious value. Nonetheless, we believe it is important to explicitly acknowledge this limitation in applications and be cautious with interpretation, and if possible, design and conduct sensitivity analysis. Alternatively, one could consider the framework developed by \cite{didelez2012direct,vanderweele2014effect} and relax sequential ignorability to allow for observed treatment-induced mediator-outcome confounding. However, this framework targets a different set of causal estimands from those considered in this paper, and thus would have to be modified accordingly. 
	
	An important extension of our method is to incorporate time-to-event outcomes, a common practice in longitudinal studies \citep{lange2012simple,vanderweele2011causal}. For example, it is of much scientific interest to extend our application to investigate the causal mechanisms among early adversity, social bonds, GC concentrations and length of lifespan. 
	A common complication in the causal mediation analysis with time-to-event outcomes and time-varying mediators is that the mediators are not well-defined for the time period in which a unit was not observed \citep{didelez2019defining,vansteelandt2019mediation}. Within our framework, which treats the time-varying observations as realizations from a process, we can bypass this problem by imputing the underlying smooth process of the mediators in an identical range for every unit.

	\section*{Acknowledgements}
	We thank Surya Tokdar, Fernando Campos, and Georgia Papadogeorgou for helpful discussions. The majority of the data represented here was supported by the National Institutes of Health and the National Science Foundation, currently through NSF IOS 1456832, and through NIH R01AG053308, R01AG053330, R01HD088558, and P01AG031719. We also thank Duke University, Princeton University, and the University of Notre Dame for financial and logistical support. For assistance and cooperation in Kenya, we are grateful to the Kenya Wildlife Service (KWS), University of Nairobi, Institute of Primate Research (IPR), National Museums of Kenya, National Environment Management Authority, and National Commission for Science, Technology, and Innovation (NACOSTI). We also thank the members of the Amboseli-Longido pastoralist communities, and the Enduimet Wildlife Management Area for their cooperation and assistance in the field. Particular thanks go to the Amboseli Baboon Project long-term field team (R.S. Mututua, S. Sayialel, J.K. Warutere, Siodi, I.L.), and to T. Wango and V. Oudu for their untiring assistance in Nairobi. The baboon project database, Babase, is expertly managed by N. Learn and J. Gordon. Database design and programming are provided by K. Pinc.  This research was approved by the IACUC at Duke University, University of Notre Dame, and Princeton University and adhered to all the laws and guidelines of Kenya. For a complete set of acknowledgments of funding sources, logistical assistance, and data collection and management, please visit \url{http://amboselibaboons.nd.edu/acknowledgements/}.
	
	\begin{supplement}[id=suppA]
		\stitle{Supplement A to ``Causal mediation analysis for sparse and irregular longitudinal data"}
		\slink[doi]{COMPLETED BY THE TYPESETTER}
		\sdatatype{.pdf}
		\sdescription{It contains the proof of Theorem \ref{T.1} (Supplement A.1), details of the Gibbs Sampler (Supplement A.2) , examples of the imputed individual process (Supplement A.3), and simulation results with $N=500, 1000$.}
	\end{supplement}

	\begin{supplement}[id=suppB]
		\stitle{Supplement B to ``Causal mediation analysis for sparse and irregular longitudinal data"}
		\slink[doi]{COMPLETED BY THE TYPESETTER}
		\sdatatype{.zip}
		\sdescription{
			We provide the dataset from the Amboseli Baboon Research Project used in this paper and the programming code, which are also available on \url{https://github.com/zengshx777/MFPCA_Codebase}.}
	\end{supplement}

	\bibliography{FPCA_Reference}
	\bibliographystyle{imsart-nameyear}

	\appendix
	
	\section*{A.1 Proof of Theorem 1}\label{appA} We provide the mathematical proof for Theorem 1. For the first part of Theorem 1, identification of total effect, for any $z\in \{0,1\}$ we have
	\begin{align*}
		E(Y_{i}^{t}|Z_{i}=z,\mathbf{X}_{i}^{t}) =E(Y_{i}^{t}(z,\mathbf{M}_{i}(z))|Z_{i}=z,\mathbf{X}_{i}^{t})
		=E(Y_{i}^{t}(z,\mathbf{M}_{i}(z))|\mathbf{X}_{i}^{t}).
	\end{align*}
	The second equality follows from Assumption 1.
	Therefore, we prove the identification of $\tau_{\TE}^{t}$,
	\begin{align*}
		\tau_{\TE}^{t}&=\int_{\mathcal{X}}  \{ E(Y_{i}^{t}(1,\mathbf{M}_{i}(1))|\mathbf{X}_{i}^{t})-E(Y_{i}^{t}(0,\mathbf{M}_{i}(0))|\mathbf{X}_{i}^{t})\}\textup{dF}_{\mathbf{X}_{i}^{t}}(\mathbf{x}^{t}),\\
		&=\int_{\mathcal{X}}  \{E(Y_{i}^{t}|Z_{i}=1,\mathbf{X}_{i}^{t}=\mathbf{x}^{t})-E(Y_{i}^{t}|Z_{i}=0,\mathbf{X}_{i}^{t}=\mathbf{x}^{t})\}\textup{dF}_{\mathbf{X}_{i}^{t}}(\mathbf{x}^{t}),
	\end{align*}
	For the second part, identification of $\tau^{t}_{\ACME}$, we make the following regularity assumptions. Suppose the potential outcomes $Y_{i}^{t}(z,\mathbf{m})$ as a function of $\mathbf{m}$ is Lipschitz continuous on $[0,T]$ with probability one. There exists $A<\infty$, $|Y_{i}^{t}(z,\mathbf{m})-Y_{i}^{t}(z,\mathbf{m'})|\leq A||\mathbf{m}-\mathbf{m'}||_{2}$, for any $z,t,\mathbf{m},\mathbf{m'}$ almost surely.
	
	For any $z,z'\in\{0,1\}$, we have
	\begin{align*}
		\int_{\mathcal{X}} \int_{R^{[0,t]}} E(Y_{i}^{t}|Z_{i}=z,\mathbf{X}_{i}^{t}=\mathbf{x}^{t}, \mathbf{M}_{i}^{t}=\mathbf{m}) \textup{dF}_{\mathbf{X}_{i}^{t}}(\mathbf{x}^{t})\times\textup{d}\{\textup{F}_{\mathbf{M}_{i}^{t}|Z_{i}=z',\mathbf{X}_{i}^{t}=\mathbf{x}^{t}}(\mathbf{m})\}\\
		=\int_{\mathcal{X}} \int_{R^{[0,t]}}E(Y_{i}^{t}(z,\mathbf{m})|Z_{i}=z,\mathbf{X}_{i}^{t}=\mathbf{x}^{t}, \mathbf{M}_{i}^{t}=\mathbf{m})\times\textup{d}\{\textup{F}_{\mathbf{M}_{i}^{t}|Z_{i}=z',\mathbf{X}_{i}^{t}=\mathbf{x}^{t}}(\mathbf{m})\}.
	\end{align*}
	For any path $\mathbf{m}$ on the time span $[0,t]$, we make a finite partition into $H$ pieces at points $\mathcal{M}_{H}=\{t_{0}=0,t_{1}=t/H,t_{2}=2t/H,\cdots,t_{H}=t\}$. Now we consider using a step function with jumps at points $\mathcal{M}_{H}$. Denote the step function as $\mathbf{m}_{H}$, which is:
	\begin{align*}
		\mathbf{m}_{H}(x)=\begin{cases} 
			\mathbf{m}(0)=m_{0} & 0\leq x< t/H, \\
			\mathbf{m}(t/H)=m_{1} & t/H\leq x<2t/H, \\
			\cdots \\ 
			\mathbf{m}((H-1)t/H)=m_{H} & (H-1)t/H\leq x\leq t. \\
		\end{cases} 
	\end{align*}
	We wish to use this step function $\mathbf{m}_{H}(x)$ to approximate function $\mathbf{m}$. First, given $\mathbf{m}$ is Lipschitz continuous, there exists $B>0$ such that $|m(x_{1})-m(x_{2})|\leq B|x_{1}-x_{2}|$. Therefore, the step functions $\mathbf{m}_{H}$ approximates the original function $\mathbf{m}$ well in the sense that,
	\begin{align*}
		||\mathbf{m}_{H}-\mathbf{m}||_{2}\leq \sum_{i=1}^{H}\frac{t}{H} B^{2}\frac{t^{2}}{H^{2}} \asymp O(H^{-2}).
	\end{align*}
	As such we can approximate the expectation over a continuous process with expectation on a vector with values on the jumps, $(m_{0},m_{1},\cdots,m_{H})$. That is,
	\begin{align*}
		\int_{\mathcal{X}} \int_{R^{[0,t]}}E(Y_{i}^{t}(z,\mathbf{m})&|Z_{i}=z,\mathbf{X}_{i}^{t}=\mathbf{x}^{t}, \mathbf{M}_{i}^{t}=\mathbf{m})\times\textup{d}\{\textup{F}_{\mathbf{M}_{i}^{t}|Z_{i}=z',\mathbf{X}_{i}^{t}=\mathbf{x}^{t}}(\mathbf{m})\}\\
		&\asymp\int_{\mathcal{X}} \int_{R^{[0,t]}}E(Y_{i}^{t}(z,\mathbf{m}_{H})|Z_{i}=z,\mathbf{X}_{i}^{t}=\mathbf{x}^{t}, \mathbf{M}_{i}^{t}=\mathbf{m}_{H})\\
		&\times\textup{d}\{\textup{F}_{\mathbf{M}_{i}^{t}|Z_{i}=z',\mathbf{X}_{i}^{t}=\mathbf{x}^{t}}(\mathbf{m}_{H})\}+O(H^{-2}).
	\end{align*}
	This equivalence follows from the regularity condition that the potential outcome $Y_{i}^{t}(z,\mathbf{m})$ as a function of $\mathbf{m}$ is continuous with the $L_{2}$ metrics of $\mathbf{m}$. As the values of steps function $\mathbf{m}_{H}$ are completely determined by the values on finite jumps, we can further reduce to,
	\begin{align*}
		\asymp\int_{\mathcal{X}} \int_{R^{H}}E(Y_{i}^{t}(z,\mathbf{m}_{H})|Z_{i}=z,\mathbf{X}_{i}^{t}=\mathbf{x}^{t},m_{0},m_{1},m_{2},\cdots m_{H})\\
		\times\textup{d}\{\textup{F}_{m_{0},m_{1},\cdots,m_{H}|Z_{i}=z',\mathbf{X}_{i}^{t}=\mathbf{x}^{t}}(m_{0},m_{1},m_{2},\cdots m_{H})\}+O(H^{-2}).
	\end{align*}
	With Assumption 1, we can show that  
	\begin{align*}
		&\textup{d}\{\textup{F}_{m_{0},m_{1},\cdots,m_{H}|Z_{i}=z',\mathbf{X}_{i}^{t}=\mathbf{x}^{t}}(m_{0},m_{1},m_{2},\cdots m_{H})\}\\
		&=\textup{d}\{\textup{F}_{m_{0}(z'),m_{1}(z'),\cdots,m_{H}(z')|\mathbf{X}_{i}^{t}=\mathbf{x}^{t}}(m_{0},m_{1},m_{2},\cdots m_{H})\},\\
		&=\textup{d}\{\textup{F}_{\mathbf{m}_{H}(z')|\mathbf{X}_{i}^{t}=\mathbf{x}^{t}}(\mathbf{m}_{H})\}.
	\end{align*}
	With a slightly abuse of notations, we use $\mathbf{m}_{H}(z)$ to denote the potential process induced by the original potential process $\mathbf{M}_{i}^{t}(z)$ and $m_{i}(z)$ to denote potential values of $\mathbf{M}_{i}^{t}(z)$ evaluated at point $x_{i}=it/H$. Also, with the Assumption 2, we can choose a large $H$ such that $t/H\leq \varepsilon$. Then we have the following conditional independence conditions,
	\begin{align*}
		Y_{i}^{0}(z,\mathbf{m}_{H}) \independent& m_{0}|Z_{i},\mathbf{X}_{i}^{t},\\
		\{Y_{i}^{t/H}(z,\mathbf{m}_{H})- Y_{i}^{0}(z,\mathbf{m}_{H})\} \independent& (m_{1}-m_{0})|Z_{i},\mathbf{X}_{i}^{t},\mathbf{m}_{H}^{0},\\
		\{Y_{i}^{2t/H}(z,\mathbf{m}_{H})- Y_{i}^{t/H}(z,\mathbf{m}_{H})\} \independent& (m_{2}-m_{1})|Z_{i},\mathbf{X}_{i}^{t},\mathbf{m}_{H}^{t/H},\\
		\cdots& \\
		\{Y_{i}^{t}(z,\mathbf{m}_{H})- Y_{i}^{t(H-1)/H}(z,\mathbf{m}_{H})\} \independent& (m_{H}-m_{H-1})|Z_{i},\mathbf{X}_{i}^{t},\mathbf{m}_{H}^{t(H-1)/H},
	\end{align*}
	where are equivalent to,
	\begin{align*}
		Y_{i}^{0}(z,\mathbf{m}_{H}) \independent& m_{0}|Z_{i},\mathbf{X}_{i}^{t},\\
		\{Y_{i}^{t/H}(z,\mathbf{m}_{H})- Y_{i}^{0}(z,\mathbf{m}_{H})\} \independent& (m_{1}-m_{0})|Z_{i},\mathbf{X}_{i}^{t},m_{0},\\
		\{Y_{i}^{2t/H}(z,\mathbf{m}_{H})- Y_{i}^{t/H}(z,\mathbf{m}_{H})\} \independent& (m_{2}-m_{1})|Z_{i},\mathbf{X}_{i}^{t},m_{0},m_{1},\\
		\cdots& \\
		\{Y_{i}^{t}(z,\mathbf{m}_{H})- Y_{i}^{t(H-1)/H}(z,\mathbf{m}_{H})\} \independent& (m_{H}-m_{H-1})|Z_{i},\mathbf{X}_{i}^{t},m_{0},m_{1}\cdots,m_{H-1},
	\end{align*}
	as the step function $m_{H}^{it/H}$ is completely determined by values $m_{0},\cdots,m_{i}$. With the above conditional independence, we have,
	\begin{align*}
		E(Y_{i}^{t}(z,\mathbf{m}_{H})&|Z_{i}=z,\mathbf{X}_{i}^{t}=\mathbf{x}^{t},m_{0},m_{1},m_{2},\cdots m_{H})\\
		&=E(Y_{i}^{t}(z,\mathbf{m}_{H})|Z_{i}=z,\mathbf{X}_{i}^{t}=\mathbf{x}^{t}).
	\end{align*}
	With similar arguments, it also equals:
	\begin{align*}
		&E(Y_{i}^{t}(z,\mathbf{m}_{H})|Z_{i}=z,\mathbf{X}_{i}^{t}=\mathbf{x}^{t})=E(Y_{i}^{t}(z,\mathbf{m}_{H})|Z_{i}=z',\mathbf{X}_{i}^{t}=\mathbf{x}^{t}),\\
		&=E(Y_{i}^{t}(z,\mathbf{m}_{H})|Z_{i}=z,\mathbf{X}_{i}^{t}=\mathbf{x}^{t},m_{0}=m_{0}(z'),\cdots m_{H}=m_{H}(z')),\\
		&=E(Y_{i}^{t}(z,\mathbf{m}_{H})|Z_{i}=z,\mathbf{X}_{i}^{t}=\mathbf{x}^{t},\mathbf{m}_{H}(z')=\mathbf{m}_{H}),\\
		&=E(Y_{i}^{t}(z,\mathbf{m}_{H})|\mathbf{X}_{i}^{t}=\mathbf{x}^{t},\mathbf{m}_{H}(z')=\mathbf{m}_{H}).
	\end{align*}
	
	As a conclusion, we have shown that,
	\begin{align*}
		&\int_{\mathcal{X}} \int_{R^{[0,t]}}E(Y_{i}^{t}(z,\mathbf{m})|Z_{i}=z,\mathbf{X}_{i}^{t}=\mathbf{x}^{t}, \mathbf{M}_{i}^{t}=\mathbf{m})
		\times\textup{d}\{\textup{F}_{\mathbf{M}_{i}^{t}|Z_{i}=z',\mathbf{X}_{i}^{t}=\mathbf{x}^{t}}(\mathbf{m})\},\\
		& \asymp    \int_{\mathcal{X}} \int_{R^{[0,t]}}E(Y_{i}^{t}(z,\mathbf{m}_{H})|\mathbf{X}_{i}^{t}=\mathbf{x}^{t},\mathbf{m}_{H}(z')=\mathbf{m}_{H})\\
		&\times \textup{d}\{\textup{F}_{\mathbf{m}_{H}(z')|\mathbf{X}_{i}^{t}=\mathbf{x}^{t}}(\mathbf{m}_{H})\}+O(H^{-2}),\\
		& \asymp    \int_{\mathcal{X}} E(Y_{i}^{t}(z,\mathbf{m}_{H}(z'))|\mathbf{X}_{i}^{t}=\mathbf{x}^{t})+O(H^{-2}),\\
		& \asymp    \int_{\mathcal{X}} E(Y_{i}^{t}(z,\mathbf{m}(z'))|\mathbf{X}_{i}^{t}=\mathbf{x}^{t})+O(H^{-2}).
	\end{align*}
	The last equivalence comes from the regularity condition of $Y_{i}^{t}(z,\mathbf{m}(z'))$ as a function of $\mathbf{m}(z')$. Let $H$ goes to infinity, we have,
	\begin{align*}
		\int_{\mathcal{X}} \int_{R^{[0,t]}} E(Y_{i}^{t}|Z_{i}=z,\mathbf{X}_{i}^{t}=\mathbf{x}^{t}, \mathbf{M}_{i}^{t}
		=\mathbf{m}) \textup{dF}_{\mathbf{X}_{i}^{t}}(\mathbf{x}^{t})\times\textup{d}\{\textup{F}_{\mathbf{M}_{i}^{t}|Z_{i}=z',\mathbf{X}_{i}^{t}=\mathbf{x}^{t}}(\mathbf{m})\}\\=\int_{\mathcal{X}} E(Y_{i}^{t}(z,\mathbf{m}(z'))|\mathbf{X}_{i}^{t}=\mathbf{x}^{t})\textup{dF}_{\mathbf{X}_{i}^{t}}(\mathbf{x}^{t}).
	\end{align*}
	With this relationship established, it is straightforward to show that,
	\begin{align*}
		\tau_{\ACME}^{t}(z)&=\int_{\mathcal{X}} \{E(Y_{i}^{t}(z,\mathbf{m}(1))|\mathbf{X}_{i}^{t}=\mathbf{x}^{t})-E(Y_{i}^{t}(z,\mathbf{m}(0))|\mathbf{X}_{i}^{t}=\mathbf{x}^{t})\}\textup{dF}_{\mathbf{X}_{i}^{t}}(\mathbf{x}^{t}),\\
		&=\int_{\mathcal{X}} \int_{R^{[0,t]}} E(Y_{i}^{t}|Z_{i}=z,\mathbf{X}_{i}^{t}=\mathbf{x}^{t}, \mathbf{M}_{i}^{t}=\mathbf{m}) \textup{dF}_{\mathbf{X}_{i}^{t}}(\mathbf{x}^{t})\times\\
		& \quad\quad \quad  \textup{d}\{\textup{F}_{\mathbf{M}_{i}^{t}|Z_{i}=1,\mathbf{X}_{i}^{t}=\mathbf{x}^{t}}(\mathbf{m})
		-\textup{F}_{\mathbf{M}_{i}^{t}|Z_{i}=0,\mathbf{X}_{i}^{t}=\mathbf{x}^{t}}(\mathbf{m})\},
	\end{align*}
	which completes the proof of Theorem 1.
	\newpage
	\section*{A.2 Gibbs Sampler}\label{appB} 
	In this section, we provide detailed descriptions on the Gibbs sampler for the model in Section 4. We only include the sampler for mediator process as the sampling procedure is essentially identical for the outcome process. For simplicity, we introduce some notations to represent vector values, $M_{i}=(M_{i1},M_{i2},\cdots,M_{in_{i}})\in \mathcal{R}^{T_{i}}$,$X_{i}=[X_{i1},X_{i2},\cdots,X_{in_{i}}]'\in \mathcal{R}^{T_{i}\times p}$, $\psi_{r}(\mathbf{t}_{i})=[\psi_{r}(t_{i1}),\cdots,\psi_{r}(t_{in_{i}})]\in\mathcal{R}^{T_{i}}$
	\begin{enumerate}
		\item \textbf{Sample the eigen function $\psi_{r}(t),r=1,2\cdots,R$.}
		\begin{itemize}
			\item (a)$\mathbf{p}_{r}|\cdots \sim N(Q_{\psi_{r}}^{-1}l_{\psi_{r}},Q_{\psi_{r}}^{-1})$ conditional on $C_{r}\psi_{r}=0$,
			\begin{align*}
				C_{r}&=[\psi_{1},\psi_{2},\cdots,\psi_{r-1},\psi_{r+1},\cdots,\psi_{R}]'B_{G}\\
				&=[\mathbf{p}_{1},\cdots,\mathbf{p}_{r-1},\mathbf{p}_{r+1},\cdots, \mathbf{p}_{R}]B_{G}'B_{G},
			\end{align*}
			where $B_{G}$ is the basis functions evaluated at a equal spaced grids on [0,1],$\{t_{1},t_{2},\cdots,t_{G}\}$, $G=50$ for example, $B_{G}=[\mathbf{b}(t_{1}),\cdots ,\mathbf{b}(t_{G})]'\in R^{G\times (L+2)}$. The corresponding mean and covariance functions are,
			\begin{gather*}
				Q_{\psi_{r}}=\frac{\sum_{i=1}^{N}B_{i}'B_{i}\zeta_{r,i}^{2}}{\sigma_{m}^{2}}+h_{k}\Omega,\\
				l_{\psi_{r}}=\frac{\sum_{i=1}^{N}B_{i}'\zeta_{i,r}(M_{i}-X_{i}\beta_{M}^{T}-\sum_{r'\neq r}^{R}\psi_{r}(\mathbf{t_{i}})\zeta_{r',i})}{\sigma_{m}^{2}}.
			\end{gather*}
			Update the $\mathbf{p}_{r}\leftarrow\mathbf{p}_{r}/\sqrt{\mathbf{p}_{r}'B_{G}'B_{G}\mathbf{p}_{r}}=\mathbf{p}_{r}/||\psi_{r}(t)||_{2}$ to ensure $||\psi_{r}(t)||_{2}=1$ and $\psi_{r}(t)=\mathbf{b}(t)\psi_{r}$ and update $\zeta_{r,i}=\rightarrow \zeta_{r,i}*||\psi_{r}(t)||_{2}$ to maintain likelihood function.
			\item (b)$h_{k}|\cdots \sim \textup{Ga}((L+1)/2,\psi_{r}'\Omega\psi_{r})$ truncated on $[\lambda_{r}^{2},10^{4}]$.
		\end{itemize}
		\item \textbf{Sample the principal score $\zeta_{r,i}$}.$\zeta_{r,i}|\cdots \sim N(\mu_{r}/\lambda_{r}^{2							},\lambda_{r}^{2})$
		\begin{gather*}
			\sigma_{r}^{2}=(||\psi_{r}(\mathbf{t}_{i})||_{2}^{2}/\sigma_{m}^{2}+\xi_{i,r}/\lambda_{r}^{2})^{-1},\\
			\mu_{r}=\frac{(M_{i}-X_{i}\beta_{M}^{T}-(\sum_{r'\neq r}\psi_{r'}(\mathbf{t}_{i})\zeta_{r',i}))'\psi_{r}(\mathbf{t_{i}}) }{\sigma_{\varepsilon}^{2}}+\frac{(\tau_{0,r}(1-Z_{i})+\tau_{1,r}Z_{i})\xi_{i,r} }{\lambda_{r}^{2}}.
		\end{gather*}
		\item \textbf{Sample the causal parameters $\chi_{0}^{r},\chi_{1}^{r}$}. Let
		$\chi_{z}=(\chi_{z}^{r},\cdots,\chi_{z}^{R}),z=0,1$,$\chi_{z}^{r}|\cdots \sim N(Q_{z,r}^{-1}l_{z,r},Q_{z,r}^{-1})$.
		\begin{gather*}
			Q_{z,r}=(\sum_{i=1}^{N} \xi_{r,i}\mathbf{1}_{Z_{i}=z}/\lambda_{r}^{2}+1/\sigma_{\chi_{r}}^{2})^{-1},\\
			l_{z,r}=\sum_{i=1}^{N} \zeta_{r,i}\xi_{r,i}\mathbf{1}_{Z_{i}=z}/\lambda_{r}^{2}.
		\end{gather*}
		\item \textbf{Sample the coefficients $\beta_{M}$}. The coefficients for covariates are $\beta_{M}|\cdots \sim N(Q_{\beta}^{-1}\mu_{\beta},Q_{\beta}^{-1})$,
		\begin{gather*}
			Q_{\beta}=X'X/\sigma_{m}^{2}+100^{2}I_{\textup{dim}(X)},\\
			\mu_{\beta}=\sum_{i=1}^{N}X_{i}'(M_{i}-\sum_{r=1}^{R}\psi_{r}(\mathbf{t_{i}})\zeta_{i,r})/\sigma_{m}^{2}.
		\end{gather*}
		\item \textbf{Sample the precision/variance parameters}.
		\begin{itemize}
			\item (a) $\sigma_{m}^{-2}|\cdots\sim \textup{Ga}(\sum_{i=1}^{N}T_{i}/2,\sum_{i=1}^{N}||M_{i}-X_{i}\beta_{M}'-\sum_{r=1}^{R}\psi_{r}(\mathbf{t_{i}})\zeta_{i,r}||_{2}^{2}/2)$
			\item (b) $\sigma_{\chi_{r}}^{2}|\cdots$,
			\begin{gather*}
				\delta_{\chi_{1}}|\cdots \sim \textup{Ga}(a_{\chi_{1}}+R,1+\frac{1}{2}\sum_{r=1}^{R}\chi_{1}^{(r)}(\chi_{0}^{r 2}+\chi_{1}^{r 2})),\chi_{l}^{(r)}=\prod_{i=l+1}^{r}\delta_{\chi_{i}}\\
				\delta_{\chi_{r}}|\cdots \sim \textup{Ga}(a_{\chi_{2}}+R+1-r,1+\frac{1}{2}\sum_{r'=r}^{R}\chi_{r'}^{(r)}(\tau_{0}^{r' 2}+\chi_{1}^{r' 2})),r\geq 2,\\
				\sigma_{\chi_{r}}^{-2}=\prod_{r'=1}^{r}\delta_{\chi_{r'}}.
			\end{gather*}
			\item (c)$\lambda_{r}^{2}|\cdots$,
			\begin{gather*}
				\delta_{1}|\cdots\sim \textup{Ga}(a_{1}+RN/2,1+\frac{1}{2}\sum_{r=1}^{R}
				\chi_{1}^{(r)'}\xi_{i,r}(\zeta_{i,r}-(1-Z_{i})\chi_{0}^{r}-Z_{i}\chi_{1}^{r})^{2},\\ \chi_{l}^{(r)'}=\prod_{i=l+1}^{r}\delta_{i},
			\end{gather*}
			\begin{align*}
				\delta_{r}|\cdots \textup{Ga}(a_{2}+&(R-r+1)N/2,\\
				&1+\frac{1}{2}\sum_{r'=r}^{R}\chi_{r'}^{(r)'}\xi_{i,r'}(\zeta_{i,r'}-(1-Z_{i})\chi_{0}^{r'}-Z_{i}\chi_{1}^{r'})^{2}),r\geq 2,\\
				&\lambda_{r}^{-2}=\prod_{r'=1}^{r}\delta_{r'}.
			\end{align*}
			\item (d) $\xi_{i,r}|\cdots\sim \textup{Ga}(\frac{v+1}{2},\frac{1}{2}(v+(\zeta_{i,r'}-(1-Z_{i})\chi_{0}^{r'}-Z_{i}\chi_{1}^{r'})^{2}/\lambda_{r}^{2}))$.
			\item (e) $a_{1},a_{2},a_{\chi_{1}},a_{\chi_{2}}$ can be sampled with Metropolis-Hasting algorithm.
		\end{itemize}
	\end{enumerate}
	The sampling for the outcomes model $Y_{ij}$ is similar to that for the mediator model except that we added the imputed value of the mediator process $M(t_{ij})$ as a covariate.
	\newpage
	
	\section*{A.3 Individual Imputed Process}\label{appC}
\begin{figure}[H]
		\centering
		\includegraphics[width=0.45\textwidth]{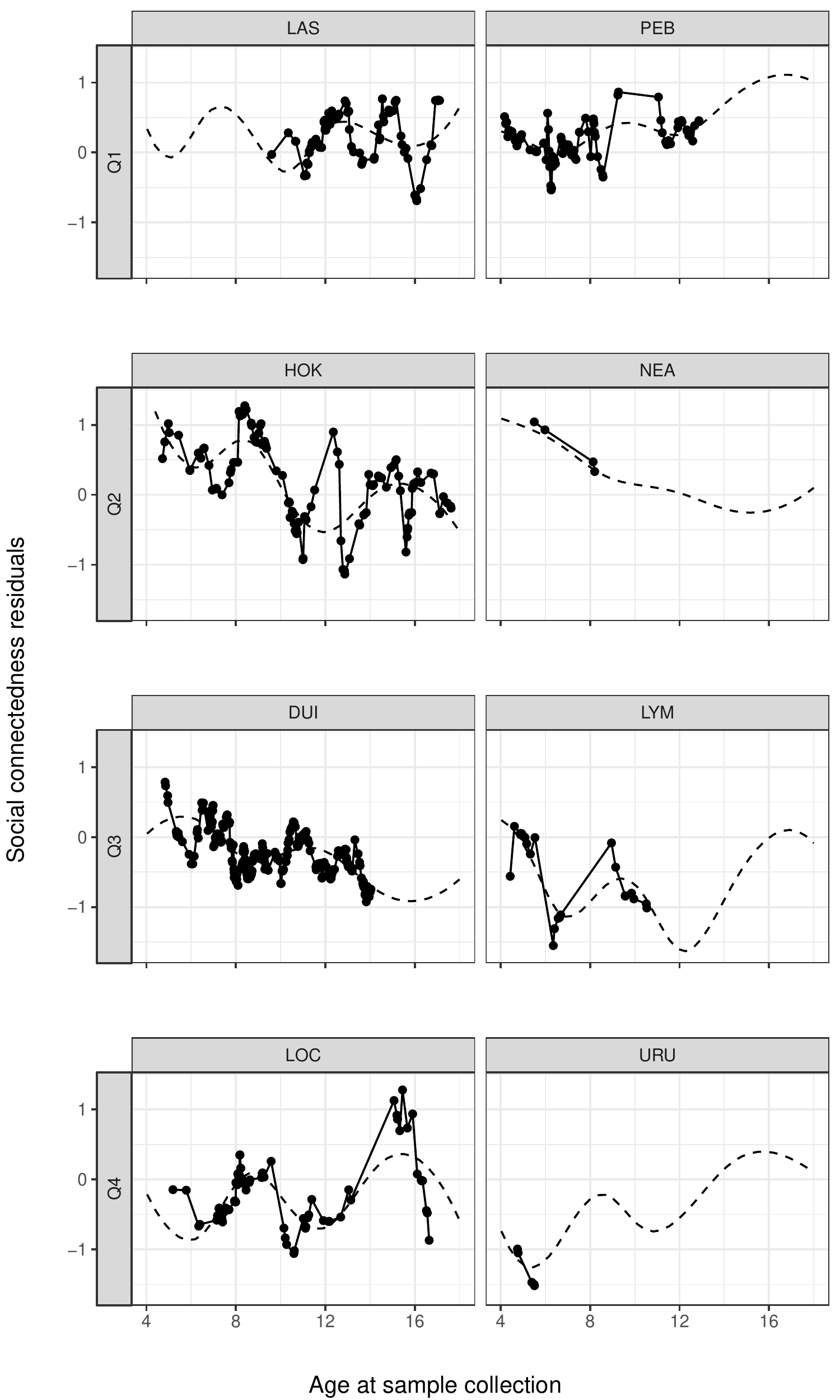}
		\includegraphics[width=0.45\textwidth]{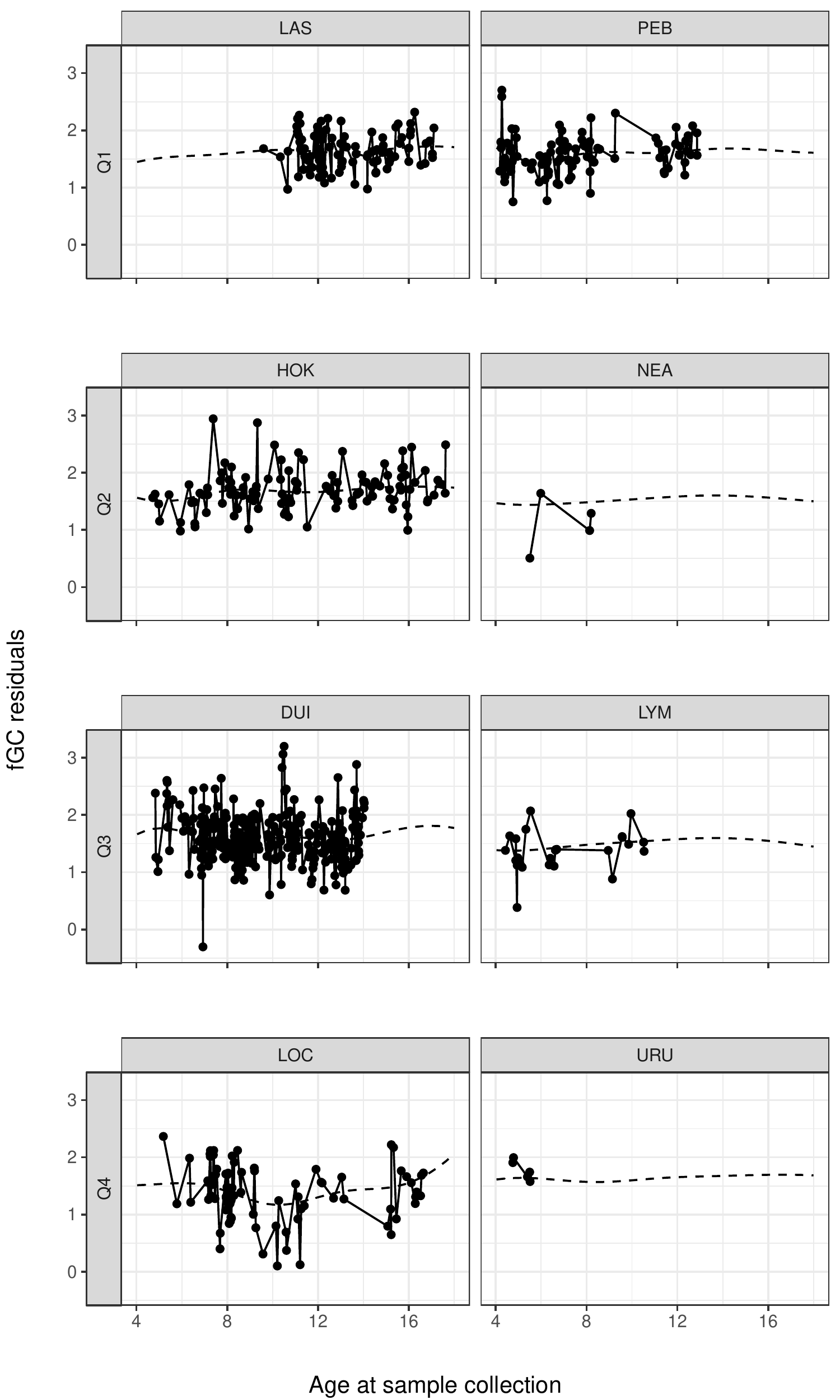}
		\caption{The imputed underlying smooth process against the observed trajectories for social bonds (left panel) and GC concentrations (right panel).}
		\label{fig:imputation}
	\end{figure}

	Figure \ref{fig:imputation} shows the posterior means of the imputed smooth processes of the mediators and the outcomes against their respective observed trajectories of eight randomly selected subjects in the sample. For social bonds (left panel of Figure \ref{fig:imputation}), the imputed smooth process adequately captures the overall time trend of each subject while reduce the noise in the observations, evident in the subjects with code name HOK, DUI and LOC.  
	
	For the subjects with few observations or observations concentrating in a short time span, such as subject NEA, the imputed process matches the trend of the observations while extrapolating to the rest of the time span with little information. FPCA achieves this by borrowing information from other units when learning the principal component on the population level. Compared with social bonds, variation of the adult GC concentrations across the lifespan is much smaller. In the right panel in Figure \ref{fig:imputation}, we can see the imputed processes for the GC concentrations are much flatter than those for social bonds. It appears that most variation in the GCs trajectories is due to noise rather than intrinsic developmental trend.
	\section*{A.4 Simulation results for sample size $N=500,1000$}
\begin{figure}[!ht]
		\centering
		\includegraphics[width=0.95\textwidth]{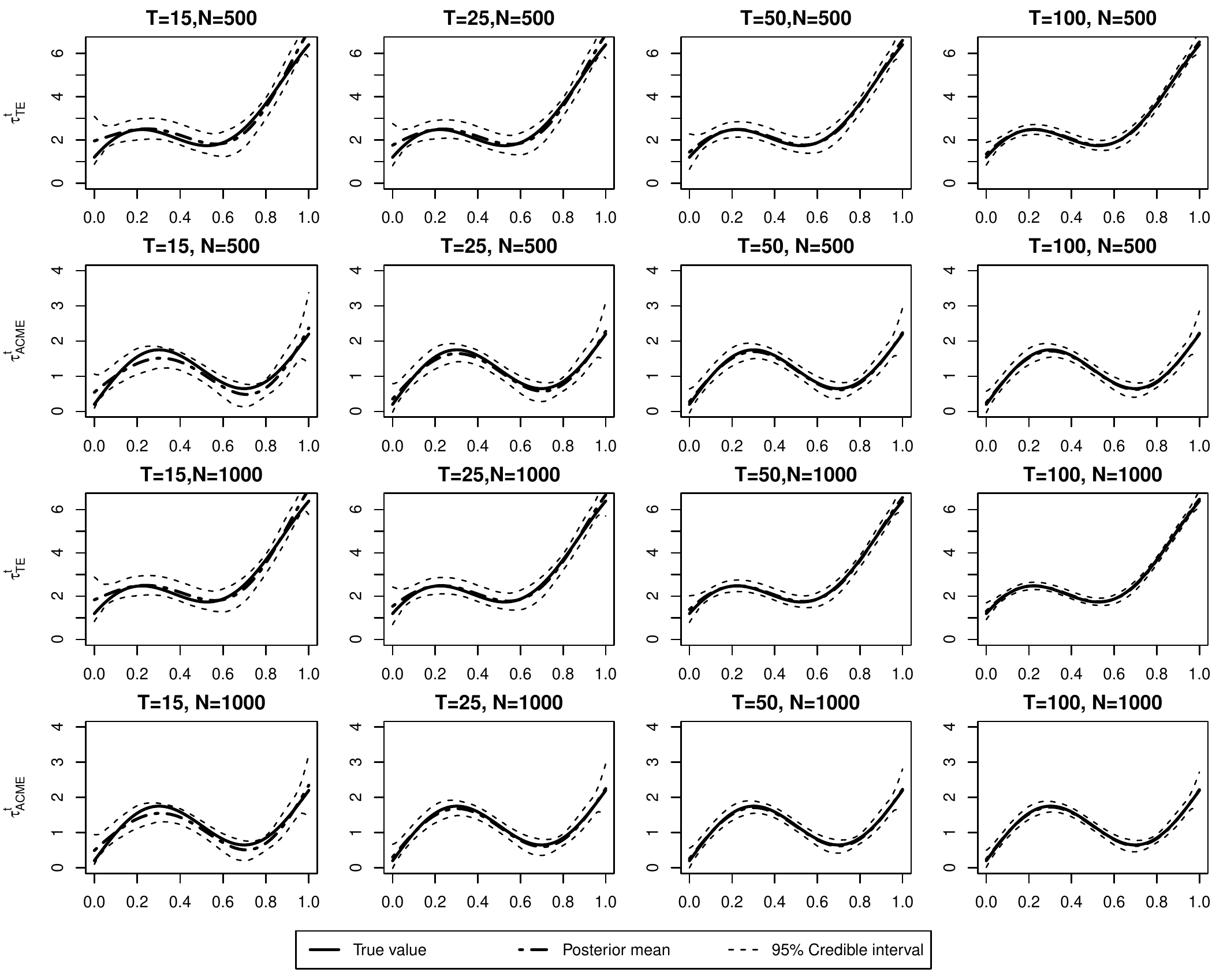}
		\caption{Posterior mean of  $\tau_{\TE}^{t}$,$\tau_{\ACME}^{t}$ and 95\% credible intervals in one simulated dataset under each level of sparsity. The top two rows are the ones when setting $N=300$ and the bottom two rows are the case when $N=500$. The solid lines are the true surfaces for $\tau_{\TE}^{t}$ and $\tau_{\ACME}^{t}$ }
		\label{fig:MFPCA_large}
	\end{figure}

	We provide the detailed simulation results on the performance of MFPCA when sample size $N$ equals $300,500$ here. In Figure \ref{fig:MFPCA_large}, we draw the posterior mean and the 95\% credible intervals for MFPCA of $\tau_{\TE}^{t}$,$\tau_{\ACME}^{t}$ across different levels of sparsity. The MFPCA produces the point estimations that are close to the true values and the credible intervals covering the true process. In Table \ref{tab:simu_comparison}, we compare the bias, RMSE and coverage rate of the proposed method with random effects and GEE approaches. Across different levels of sparsity, the MFPCA shows a lower bias and the RMSE compared with the other methods. Also, the coverage rate of the MFPCA for $\tau_{\TE}^{t}$,$\tau_{\ACME}^{t}$ becomes close to the nominal level 95\% when the sample size $N$ and the observations per unit $T$ is larger.

	\begin{table}[!ht]
		\centering
		\caption{Absolute bias, RMSE and coverage rate of the 95\% confidence interval of MFPCA, the random effects model and the generalized estimating equation (GEE) model under different sparsity levels with $N=500,1000$. }
		\label{tab:simu_comparison}
		\begin{tabular}{ccccccc}
			\hline
			&\multicolumn{3}{c}{$\tau_{\TE}$} &\multicolumn{3}{c}{$\tau_{\ACME}$}\\
			Method &Bias & RMSE & Coverage&Bias & RMSE & Coverage\\
			\hline
			&\multicolumn{6}{c}{$T$=15, $N$=500}\\
			MFPCA&0.079 &0.109 &91.4\%& 0.104&0.196& 90.8\% \\
			Random effects&0.103&0.132& 87.4\% &0.424&0.967&84.2\%\\
			GEE &0.117&0.163 & 87.1\%&0.531&1.421& 79.5\%\\
			\hline
			&\multicolumn{6}{c}{$T$=25, $N$=500}\\
			MFPCA&0.076 &0.102 &93.8\%& 0.096&0.189& 92.0\% \\
			Random effects&0.092&0.118& 91.5\% &0.397&0.894&88.4\%\\
			GEE &0.095&0.134 & 90.7\%&0.523&1.032& 89.1\%\\
			\hline
			&\multicolumn{6}{c}{$T$=50, $N$=500}\\
			MFPCA&0.061 &0.095 &94.3\%& 0.073&0.176& 92.4\% \\
			Random effects&0.068&0.097& 93.7\% &0.078&0.170&92.8\%\\
			GEE &0.073&0.104 & 93.2\%&0.089&0.323& 92.5\%\\
			\hline
			&\multicolumn{6}{c}{$T$=100, $N$=500}\\
			MFPCA&0.035 &0.068 &95.2\%& 0.046&0.123& 94.8\% \\
			Random effects&0.033&0.061& 95.1\% &0.045&0.129&94.3\%\\
			GEE &0.035&0.067 & 94.5\%&0.051&0.130& 94.2\%\\
			\hline
			&\multicolumn{6}{c}{$T$=15, $N$=1000}\\
			MFPCA&0.065 &0.097 &91.9\%& 0.097&0.164& 91.3\% \\
			Random effects&0.085&0.123& 90.7\% &0.387&0.885&89.6\%\\
			GEE &0.093&0.145 & 90.4\%&0.489&1.103& 84.6\%\\
			\hline
			&\multicolumn{6}{c}{$T$=25, $N$=1000}\\
			MFPCA&0.053 &0.081 &93.9\%& 0.088&0.185& 93.1\% \\
			Random effects&0.058&0.097& 93.0\% &0.230&0.526&91.7\%\\
			GEE &0.068&0.115 & 92.4\%&0.297&0.611& 91.5\%\\
			\hline
			&\multicolumn{6}{c}{$T$=50, $N$=1000}\\
			MFPCA&0.036 &0.063 &94.2\%& 0.054&0.221& 93.6\% \\
			Random effects&0.035&0.068& 94.2\% &0.052&0.203&93.2\%\\
			GEE &0.040&0.073 & 93.9\%&0.057&0.214& 93.1\%\\
			\hline
			&\multicolumn{6}{c}{$T$=100, $N$=1000}\\
			MFPCA&0.020 &0.053 &95.1\%& 0.023&0.087& 94.8\% \\
			Random effects&0.023&0.050& 94.8\% &0.021&0.097&94.4\%\\
			GEE &0.021&0.054 & 94.6\%&0.027&0.098& 94.5\%\\
			\hline
		\end{tabular}
	\end{table}

\end{document}